\documentclass[%
 reprint,
 amsmath,amssymb,
 aps,
]{revtex4-2}

\usepackage{graphicx}
\usepackage{dcolumn}
\usepackage{bm}
\usepackage[T1]{fontenc}
\usepackage[utf8]{inputenc}

\begin{document}

\preprint{APS/123-QED}

\title{Two-electron spectrum of a silicon quantum dot}

\author{Bilal Tariq}
\email{bilaltariq.phy@gmail.com}
\affiliation{%
Department of Physics, University at Buffalo, SUNY, Buffalo, New York 14260-1500, USA \\
National Center for Physics, Quaid-i-Azam University Campus, Islamabad 44000, Pakistan
}
\author{Xuedong Hu}%
 \email{xhu@buffalo.edu}
\affiliation{%
Department of Physics, University at Buffalo, SUNY, Buffalo, New York 14260-1500, USA.
}%

\date{\today}

\begin{abstract}
The energy spectrum and wave functions of electrons in a single silicon quantum dot provide valuable insights into the capabilities and limitations of such a system in quantum information processing. Here we investigate the low-lying singlet and triplet configurations and spectra in a two-electron silicon quantum dot. To build toward a comprehensive understanding, we first examine the competition between Coulomb interaction and electron kinetic and confinement energy in the absence of valley-orbit coupling, as well as consequences of valley blockade in the presence of an ideal smooth interface. For realistic interfaces the variations in the magnitude and phase of valley-orbit coupling lead to inter-valley leakage, particularly when orbital splittings approach the valley splitting. In our study we particularly focus on the impact on the compositions of low-lying singlets and triplets. We find that for experimentally relevant parameter regimes the ground singlet and triplet states usually contain multiple configurations with significant weights as a result of a complicated competition among valley-orbit coupling, confinement potential, and Coulomb interaction. We further analyze the effects of an out-of-plane magnetic field on these the two-electron spectra. Our findings could have important implications for spin qubits in Si quantum dot in various contexts, such as qubit encoding and spin measurement.
\end{abstract}

\flushbottom
\maketitle

\thispagestyle{empty}

\section{Introduction}
The excellent spin coherence and strong exchange interaction \cite{itoh2014isotope, veldhorst2014addressable, muhonen2014storing, borjans2019single}, as illustrated by recent experimental demonstration of high-fidelity single-qubit and two-qubit gates  \cite{yoneda2018quantum, yang2019silicon, mills2022two, noiri2022fast, xue2022quantum, philips2022universal, lawrie2023simultaneous, neyens2024probing}, make spin qubits in Si promising building blocks for a scalable universal quantum computer \cite{loss1998quantum, kane1998silicon, hanson2007spins, zwanenburg2013silicon, burkard2023semiconductor}.
A key challenge for electron spin qubits in silicon is the valley degree of freedom, originating from the six-fold degeneracy of the conduction band minimum in bulk Si.  While this degeneracy is usually completely lifted in gated Si quantum dots by the host heterostructure and the interface \cite{koiller2001exchange, hada2004exchange, friesen2006magnetic, friesen2007valley, goswami2007controllable, shi2013spin, zwanenburg2013silicon, benito2019optimized, huang2021electric}, the first valley excited state is often only tens of $\mu$eV above the ground orbital state (the valley splitting), posing an ever present channel for qubit leakage.  Indeed, both magnitude (i.e. the valley splitting, which detemines the leakage gap) and phase (strong influence on tunnel coupling between neighboring quantum dots) of the valley-orbit coupling play significant roles in spin qubit control and coherence \cite{yang2013spin, huang2014spin, borjans2019single, hollmann2020large, burkard2016dispersive, gamble2016valley, zhao2018coherent, tagliaferri2018impact, tariq2019, ferdous2018valley, hosseinkhani2020electromagnetic, voisin2020valley, borjans2021probing, ferdous2018interface, jock2018silicon, tanttu2019controlling, cifuentes2024bounds}, and both are strongly affected by
disorders at the interface and in the alloy barrier \cite{li2010exchange, culcer2010quantum, culcer2010interface, friesen2010theory, friesen2007valley, borselli2011pauli, hollmann2020large, zajac2015reconfigurable}.  While improvement in material growth and device fabrication has led to better control of valley physics in recent years, it will remain a formidable technical challenge for coherent control of spin qubits in Si in the foreseeable future and has to be accounted for carefully. 
Another remaining technical challenge for spin qubits is their measurement.  With magnetic interactions too weak, current approaches are all based on spin-charge conversion, where spin information is mirrored in a certain charge property, 
which is then measured electrically \cite{elzerman2004single, petta2005coherent, mi2018coherent}.  While great progress has been made in measuring single spins, the measurement speed is still relatively slow in comparison with spin decoherence. With qubit measurement an essential ingredient to quantum error correction, faster spin measurement with high fidelity will be key for spin qubits toward the long-term goal of building a universal quantum computer.

Some spin measurement approaches are based on Pauli spin blockade (PSB) \cite{ono2002current, johnson2005singlet, petta2005coherent, lundberg2024non}, which relies on the spectral difference between a two-electron double dot and a two-electron single dot to differentiate a spin singlet from a triplet state.  To establish the efficacy of PSB in enabling fast and high-fidelity spin measurement, one needs to first clarify the two-electron energy spectrum and state composition in both a single and a double quantum dot.  In this context, valley-orbit coupling, as well as spin-valley coupling, could introduce additional dynamics and narrow the effective range of PSB \cite{jones2019spin, blumoff2022fast, anderson2022high}.  

In this work, we calculate the two-electron spectrum of a single quantum dot in the presence of valley-orbit coupling using the configuration interaction (CI) approach, with particular interest in the impact of valley physics on the contributing configurations in low-energy states.  In our calculation, we account for the fact that different orbital states may have different valley-orbit couplings due to their individual charge density distributions.  Such variations in valley-orbit coupling, particularly in its phase, break all valley-related selection rules, allowing the two-electron eigenstates to have more diverse contributions from various configurations, which could in turn have significant impact when evaluating quantities such as inter-dot tunnel coupling and spin relaxation.  

This paper is organized as follows: The next section outlines the theoretical framework of our study, including the effective mass Hamiltonians we employ for a Si quantum dot, as well as the single-electron orbitals and two-electron orbitals for our CI calculation.  
The results on ground state exchange splitting and state compositions of the singlet and triplet states in the absence of valleys (or the limit of infinite valley splitting) are provided in Section \ref{sec3}. In Section \ref{sec4}, we explore the effects of a uniform valley-orbit coupling (such as for a smooth interface) on the two-electron spectrum, and clarify how the energy levels of the singlet and triplet states are modified. The theoretical model of the valley-orbit coupling for a realistic interface (such as an interface with atomic scale steps) is given in Section \ref{sec5}, where we further clarify valley-orbit couplings for different orbital states. In Section \ref{sec6} we present the full energy spectrum of a realistic two-electron Si quantum dot, as well as the state compositions of the low-lying singlet and triplet states. In Section \ref{sec7}, we investigate the impact of the magnetic field on the orbital energy levels with and without valley. Our conclusions, with a summary of our findings, are presented in Section \ref{sec8}.

\section{Theoretical Model}\label{sec2}

To solve the two-electron problem in a Si quantum dot (QD), we start by constructing the single-electron basis within the effective mass approximation.  For a gated two-dimensional QD, the confinement potential along the growth ($z$) direction (i.e. the (001) direction for the Si lattice) is much stronger than the in-plane ($xy$) directions.  Given a magnetic field along the growth direction, the total effective-mass Hamiltonian within each conduction band valley can be separated into the in-plane and out-of-plane parts as 
\begin{eqnarray}
    H_0 & = & H_{xy} + H_z \notag\\ 
    H_{xy}&=&\frac{-\hbar^2}{2m_t}\left[\left(\partial_x-e A_x\right)^2 +\left(\partial_y-e A_y\right)^2\right]+\frac{\hbar^2 \left(x^2+y^2\right)}{2m_t \ell_0^4}\notag\\
    H_{z}&=&\frac{-\hbar^2\left(\partial_z^2\right)}{2m_l}+\frac{\hbar^2 z^2}{2m_l b^4}\notag
\end{eqnarray}
where $\boldsymbol{A}=B/2\left(-y,x,0\right)$ is the vector potential in the symmetric gauge, $m_t = 0.192 m_0$ and $m_l = 0.916 m_0$ are the effective masses of the electron ($m_0$ being the bare electron mass) along the transverse and longitudinal directions in Si and $\ell_0$ and $b$ are the in-plane and out of plane effective radii of the QD when $B=0$. The in-plane confinement energy is related to the confinement radius as $E_0=\frac{\hbar^2}{m_t\ell_0^2}$. In our calculations we typically use confinement energies of $1.0$ and $0.5$ meV (with corresponding effective radius of the QD at 20 and 28 nm) respectively to represent a small and a large QD.  We choose a harmonic confinement along the growth $z$ direction for simplicity in the eventual two-electron calculation.  With valley physics included phenomenologically in the current study, this choice of confinement should not lead to any loss of generality. The solutions to Hamiltonian $H_0$ are well known: the eigenstates are products of the so-called Fock-Darwin states in the $xy$ direction and simple harmonic oscillator states in the $z$ direction.

With the Si/SiGe heterostructure grown along the (001) direction of the Si lattice, only the $z$ and $\bar{z}$ valleys are involved in the low-energy spectrum for the confined electrons, making $m_t$ and $m_l$ the in-plane and out-of-plane effective masses for them.  Within the effective mass approximation, the single-electron wave function can be expanded on a basis set of products of an envelope function and a Bloch state at the $z$ and $\bar{z}$ minima of the conduction band:  
\begin{equation}
    D^i_{\xi}(\textbf{r})=\mathcal{F}^i(\textbf{r}) u_{\xi}(\textbf{r})e^{ -i\xi k_0z} \,.
\end{equation}
Here $k_0=0.85(2\pi/a_0)$ is the location of the valley minima along the $z$ direction in the first Brillouin Zone, and $u_{\xi}(\textbf{r})$ are the periodic part of the corresponding  Bloch state. Where, $\xi=\{\pm 1\}$ corresponds to the $z$ and $\bar{z}$ valley states. The envelope function $\mathcal{F}$ can be decomposed into a product of the in-plane and out-of-plane parts:
\[\mathcal{F}^i(\boldsymbol{r})=\psi^i(x,y)\phi_0(z),\]
where the superscript $i\in (n,l)$ represents the in-plane orbital quantum numbers: $\psi^i(x,y)\equiv \psi_{nl}(r,\theta)$. With the magnetic field along the growth direction, the in-plane basis states are the Fock-Darwin states, while for the out-of-plane direction $\phi_0(z)$ is the simple harmonic ground state.  In this study we assume no excitation along $z$ direction due to the typically much stronger confinement.  More explicitly,
    \begin{eqnarray}
    \psi_{nl}(r,\theta)&=&\sqrt{\frac{n!}{\pi \ell_B^2\left(n+|l|\right)!}}\left(\frac{r}{\ell_B}\right)^{|l|} e^{-il\theta-\frac{r^2}{2 \ell_B^2}} L_n^{|l|}\left(\frac{r^2}{\ell_B^2}\right) \,, \notag\\
    \phi_0(z) &=& \frac{1}{\sqrt{\pi} b}\; e^{-\frac{z^2}{2 b^2}} \,.\notag
    \end{eqnarray}
Here $L_n^{|l|}(x)$ are the associated-Laguere polynomials, and $\ell_B$ is the magnetic confinement length defined as \cite{hu2000hilbert, burkard1999coupled}
\[\ell_B=\ell_0 \left[1+\frac{e^2B^2\ell_0^4}{4\hbar^2}\right]^{-\frac{1}{4}} \,.\] 

Given the single-electron basis states $D^i_{\xi}(\textbf{r})$, we perform a configurational Interaction (CI) calculation to obtain the spectrum and eigenstates of two electrons in a single Si QD.  The effective mass Hamiltonian of the two electrons is given by:
\begin{equation}
    H(\boldsymbol{r}_1,\boldsymbol{r}_2)= H_0(\boldsymbol{r}_1)+H_0(\boldsymbol{r}_2)+H_c(\boldsymbol{r}_1-\boldsymbol{r}_2) \,,
\end{equation}
where $H_0$ is the single electron Hamiltonian discussed above, and $H_c=\frac{1}{4\pi\epsilon_0\epsilon_r|\boldsymbol{r}_1-\boldsymbol{r}_2|}$ is the electron-electron Coulomb interaction, with an electrostatic dielectric constant $\epsilon_r=11.8$ for conduction electrons in Si. 

We do not include spin-orbit coupling in this study, so that electron spin is a good quantum number, which allows us to divide the two-electron states into spin singlet and triplet sectors, with the orbital part of the two-electron basis states being the symmetric and anti-symmetric combinations of the single-electron orbitals,
\begin{eqnarray}\label{SY0}
     \Psi_{\xi  \xi,S} ^{ii}(\boldsymbol{r}_1,\boldsymbol{r}_2)&=&D_\xi ^i(\boldsymbol{r}_1) D_\xi ^i(\boldsymbol{r}_2)\notag\\
       \Psi_{\xi  \xi',S} ^{ij}(\boldsymbol{r}_1,\boldsymbol{r}_2)&=&
  \frac{1}{\sqrt{2}}\left[D_\xi^i(\boldsymbol{r}_1) D_{\xi'}^j(\boldsymbol{r}_2) + D_{\xi'}^j(\boldsymbol{r}_1) D_\xi ^i(\boldsymbol{r}_2) \right] \notag\\
 \end{eqnarray}
  \begin{eqnarray}\label{AS0}
     \Psi_{\xi\xi',T}^{ij}(\boldsymbol{r}_1,\boldsymbol{r}_2)= \frac{1}{\sqrt{2}}\left[ D_\xi^i(\boldsymbol{r}_1) D_{\xi'}^j(\boldsymbol{r}_2) - D_{\xi'}^j(\boldsymbol{r}_1) D_\xi^i(\boldsymbol{r}_2) \right]\notag\\
 \end{eqnarray}
In the above equations, $\Psi_{\xi  \xi,S} ^{ii}(\boldsymbol{r}_1,\boldsymbol{r}_2)$ represents the case where both electrons occupy the same orbital and valley states. For all other configurations, we obtain the singlet and the corresponding triplet combinations. Due to the need to consider valley-orbit coupling, we have included the Bloch states here in our two-electron basis states.  Due to their orthogonality and fast-oscillating nature, the Bloch states do not contribute to the computation of matrix elements other than providing selection rules.  
 
The number of two-electron basis states grows rapidly as we increase the number of single-electron orbitals. Our minimum calculations, in the absence of out-of-plane excitations and valley excitation, are built on $SPD$ (or 6) in-plane orbitals.  There are correspondingly 15 triplet and 21 singlet configurations. Our largest calculations are built on $SPDFG$ states, with totally 15 single-electron orbitals, when the two-electron basis set increases to 105 triplet and 120 singlet states. We expand the two-electron Hamiltonian on these basis sets, then diagonalize the Hamiltonian numerically. 
In the literature there exist studies using much larger basis sets \cite{friesen2010theory, barnes2011screening, shim2018barrier, anderson2022high, foulk2024theory, rodriguez2025dressed}, with the goal of obtaining accurate multi-electron spectrum.  Our main focus in this study is instead on state compositions of the low-lying states of a two-electron Si quantum dot, for which purpose the $SPDFG$ basis set turns out to be more than sufficient, as we demonstrate in the next section \cite{shim2018barrier, anderson2022high}.
%
%

When valley degree of freedom is included, the number of single-electron orbitals doubles. The minimum number of triplet and singlet basis states thus increase to 66 and 78 accordingly, and the maximum increase to 435 and 465, respectively, in our calculations.
 
\begin{figure}[t]
\includegraphics[width=0.8\linewidth]{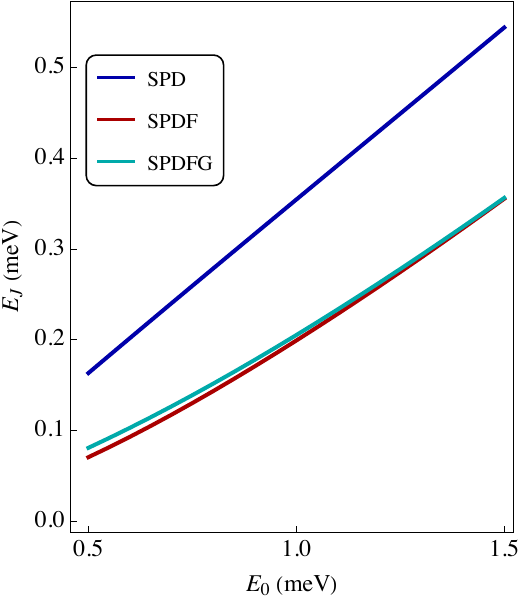}
	\caption{\textbf{Orbital exchange splitting:} (Color online) The effect of the confinement energy on the exchange splitting in the two electrons silicon quantum dot. The valleys contributions are not included here and each curve represents the different numbers of CI basis sets included in the calculation.
	}\label{fig_1}
\end{figure}

\section{Orbital Exchange Energy and State Composition}\label{sec3}

In this Section we first evaluate two-electron exchange splitting in a single dot when valley splitting is much larger than the orbital excitation energy from the lateral confinement in a quantum dot with ideal interfaces.  At this limit, all the low-energy singlet and triplet states share the same Bloch state, such that it does not influence in any way the low-energy spectrum of the two-electron quantum dot.  The two-electron basis states for our CI calculation can thus be expressed simply in terms of the envelope functions:
 \begin{eqnarray}\label{SY}
     \Psi_S^{ij}(\boldsymbol{r}_1,\boldsymbol{r}_2)=
     \begin{cases}
  \mathcal{F}^i(\boldsymbol{r}_1)\mathcal{F}^i(\boldsymbol{r}_2) & \text{$i=j$ } \\
  \frac{1}{\sqrt{2}}\left[\mathcal{F}^i(\boldsymbol{r}_1)\mathcal{F}^j(\boldsymbol{r}_2)+\mathcal{F}^j(\boldsymbol{r}_1)\mathcal{F}^i(\boldsymbol{r}_2) \right] & \text{$i\neq j$}
\end{cases}\notag\\
 \end{eqnarray}
  \begin{eqnarray}\label{AS}
     \Psi_T^{ij}(\boldsymbol{r}_1,\boldsymbol{r}_2)= \frac{1}{\sqrt{2}}\left[\mathcal{F}^i(\boldsymbol{r}_1)\mathcal{F}^j(\boldsymbol{r}_2)-\mathcal{F}^j(\boldsymbol{r}_1)\mathcal{F}^i(\boldsymbol{r}_2) \right]. \notag\\
 \end{eqnarray}
From here on, when valley degree of freedom is neglected, we call $\Psi_S^{ij}$ the $ij$ configuration, with the two single-electron orbitals $\mathcal{F}^i$ and $\mathcal{F}^j$ states.

An arbitrary two-electron wave function can in general be expressed as a linear superposition of the two-electron basis set:
\begin{equation}
    \Phi_k\left(\boldsymbol{r}_1,\boldsymbol{r}_2\right)=\sum_{ij} c^{ij} \Psi_k^{ij}\left(\boldsymbol{r}_1,\boldsymbol{r}_2\right) \text{   for $k\in (S,T)$ }\notag 
\end{equation}
Here $S$ and $T$ refer to the singlet and triplet two-electron basis states given in Eqs.~(\ref{SY}) and (\ref{AS}), respectively.  Our CI calculation is now a matrix diagonalization problem that can be solved numerically.  With $|c^{ij}|^2$ the probability of a two-electron basis state, it is of particular interest to us as it signifies the contribution from a particular configuration in an eigenstate, and determines properties such as tunnel coupling and hyperfine coupling for the electrons. A key goal of the present project is to clarify how $|c^{ij}|^2$ depends on the various system parameters, especially the orbital-dependent valley-orbit coupling.

Our first task here is to check convergence, where we perform our CI calculation with progressively larger basis set.  We start by including up to $D$ single-electron orbitals, using in total 6 single-electron orbital states, with the corresponding 15 triplet and 21 singlet basis states. We then expand our basis by including $F$ and then $G$ states, such that the number of single-electron orbitals is increased to 10 and 15, respectively, and the number of two-electron basis states increases accordingly.

Figure \ref{fig_1} shows that in our chosen parameter regime the calculated exchange splitting between the ground singlet and triplet states is already converging when we include up to $F$ single-electron orbitals, as there is only a small quantitative correction when adding the $G$ orbitals in the calculation. The significant shift in the $SPDF$ and $SPDFG$ results as compared with the $SPD$ results indicates that important contributions from excited orbitals have been missed with the smaller basis set. 
The difference becomes smaller for stronger confinement, when the exchange energy becomes dominated by the orbital energy rather than the Coulomb interaction. To investigate these features further, we perform a detailed analysis of the state contributions to the singlet and triplet states.

\begin{figure}[t]
\includegraphics[width=\linewidth]{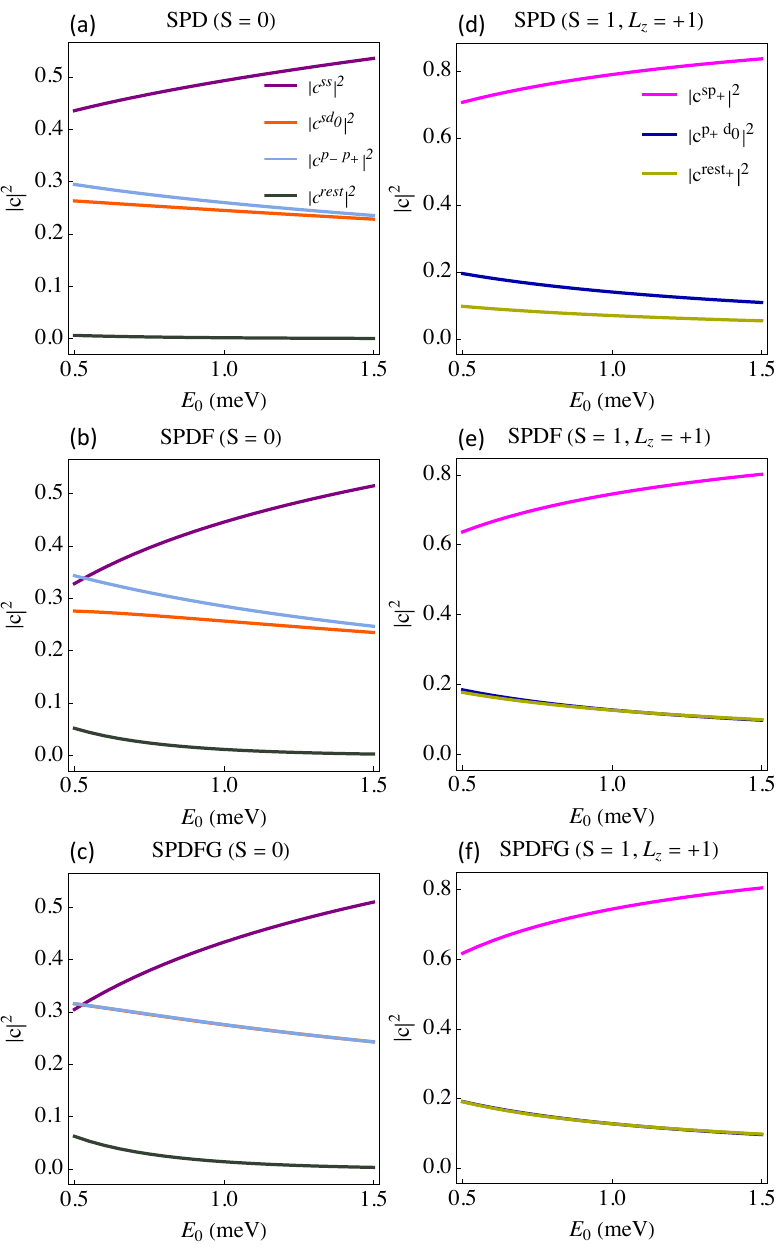}
   \caption{\textbf{Orbital state compositions:} (Color online) The dependence of state composition on the confinement energy for ground singlets (panels (a)-(c)) and triplets (panels (d)-(f)) is shown with an increasing number of excited orbital states in the calculations, excluding the valley states. The contributions from $|c^{sd_0}|^2$ and $|c^{p_-p_+}|^2$ are equal in panel (c). For triplets, the contributions from the $S=1$ and $S=-1$ orbitals are equal and thus represent as a single line. In panels (e) and (f), the contributions of $|c^{sp_-}|^2$ ($|c^{sp_+}|^2$) are approximately equal to $|c^{rest_-}|^2$ ($|c^{rest_+}|^2$).}\label{fig_2}
\end{figure}

Conventionally, it is believed that the ground singlet state in a two-electron Si quantum dot is dominated by the configuration of doubly occupied single-electron ground state, while the ground triplet is dominated by the configuration where one electron is in the ground orbital state, while the other in the first excited orbital.  To verify this belief and to clarify the features shown in Fig.~\ref{fig_1}, we plot the state composition of the ground singlet and triplet in Fig.~\ref{fig_2}. Each panel displays the results with the most important contributions to the ground singlet and triplet states as we vary the number of single-electron orbitals in the CI calculation. The sum of the remaining states' contribution is represented as $|c^{rest}|^2$. 
%
%
%
As shown in all the panels on the ground singlet state, the configuration of doubly occupied $s$-orbital usually makes the largest contribution to the ground singlet, especially for larger confinement energy, as expected. However, it also clearly does not dominate the ground singlet.  Its contribution is only slightly more than 50\% even at the larger confinement energy of 1.5 meV. As the confinement energy of the dot decreases, its contribution decreases as well. This decline is more pronounced for the cases with a larger basis set (when $F$ and $G$ orbitals are included). Indeed, for a confinement energy $E_0 = 0.5$, the contribution drops to slightly below 30\% (panel (c)), while the largest contributions now come from $sd_0$ and $p_-p_+$ configurations, each with slightly more than 30\% contribution to the ground singlet state (note that in panel (c), the contributions from $sd_0$  and $p_-p_+$ configurations are equal, resulting in a single curve.). The rest of the singlet states $|c^{rest}|^2$ have negligible contributions. 

The most prominent feature of the singlet state composition is the apparent importance of the $sd_0$ and $p_-p_+$ configurations.  A closer inspection of these two contributions as a function of increasing the size of the basis set reveals an interesting twist: From panel (a) to (c), the basis set increases in size. In panel (a) of Fig.~\ref{fig_2}, the contributions from $sd_0$ and $p_-p_+$ are somewhat different.  In panel (b), the difference increases.  However, in panel (c), which has the largest basis set, the difference shrinks to a negligible level. This trend reversal can be explained by examining the states that could couple to these two configurations, which include $\{sg_0, p_-f_+, p_+f_-, d_{--}d_{++}, d_0d_0\}$. The absence of any configuration in the CI calculation, such as when we only include up to $D$ or $F$ orbitals, leads to incomplete couplings and a misleading trend. However, when orbitals up to $G$ are all included, all states that directly couple to $sd_0$ and $p_-p_+$ configurations are accounted for (thus all lowest-order corrections are now included), which results in equal contributions from these two configurations to the ground singlet state.

The ground triplet state composition is degenerate for the orbital angular momentum $L_z = \pm 1$ because we consider an isotropic confinement potential in-plane. This symmetry breaks down with random interface roughness or an applied magnetic field along the growth direction, as will be discussed in later sections, or if the dot potential is anisotropic. As shown in Fig.~\ref{fig_2}, for $L_z = +1$ ($-1$) the major contribution comes from $sp_+$ ($sp_-$) configuration. It counts for approximately 80\% of the total probability at larger confinement energies. The next major contributions come from the $p_-d_{++}$ ($p_+d_{--}$) configuration. Contributions from the remaining states become comparable to the second dominant term in the triplet when the basis set is larger ($SPDF$ and $SPDFG$ cases). In other words, a basis set including only up to $D$ orbitals would miss some important contributions from the linear combination of lower and excited orbitals, even at smaller sizes of the quantum dot. These omissions significantly impact the convergence of the triplet at larger confinement energies, indicating that we need to include orbital states up to at least the $F$ level, similar to the case of ground singlet state.  

The results in Fig.~\ref{fig_2} show that for a typical two-electron Si quantum dot with orbital excitation energy $\lesssim 1$ meV, {\it in the absence of valleys}, we need to include excited orbital states in order to have a faithful description of the low-energy spectrum. In particular, for both the ground singlet and triplet states, orbital states up to $F$ have to be included in the basis set, and for excited singlet states, $G$ orbitals are also needed. While the doubly occupied $ss$ configuration often makes the most important contribution to the ground singlet, it is far from dominant.  Indeed, for smaller confinement energy, other configurations could be more important than the $ss$ configuration. 
%
Although a complete description would require including basis states up to the $G$ orbital, 
their inclusion only slightly changes the numerical values but does not affect the overall qualitative dependence on the confinement potential. Therefore, in the subsequent sections, where the effects of the interface step and valley-orbit coupling are discussed, we use basis states only up to the $F$ orbital, since they capture all the essential physical features.

The results of this Section provide a foundational background for incorporating valley degrees of freedom into each orbital. Next we examine the energy level shifts when introducing a constant valley splitting in each orbital state.

\section{Exchange Energy with a Constant Valley-Orbit Coupling}\label{sec4}

In the previous Section we have assumed an infinite valley-orbit coupling in order to focus on how orbital excitations contribute to the two-electron low-energy spectrum. In this section, we include the valley-orbit coupling in the simplest form, by assuming that valley-orbit coupling is a finite constant everywhere, and for all orbital excitations.  This situation represents a perfect interface without any roughness or disorder, and there is no alloy disorder in the SiGe barrier.  The magnitude of valley-orbit coupling is treated as a free parameter, with a value between 0 and 0.1 meV.  
\begin{figure}[t]
\includegraphics[width=\linewidth]{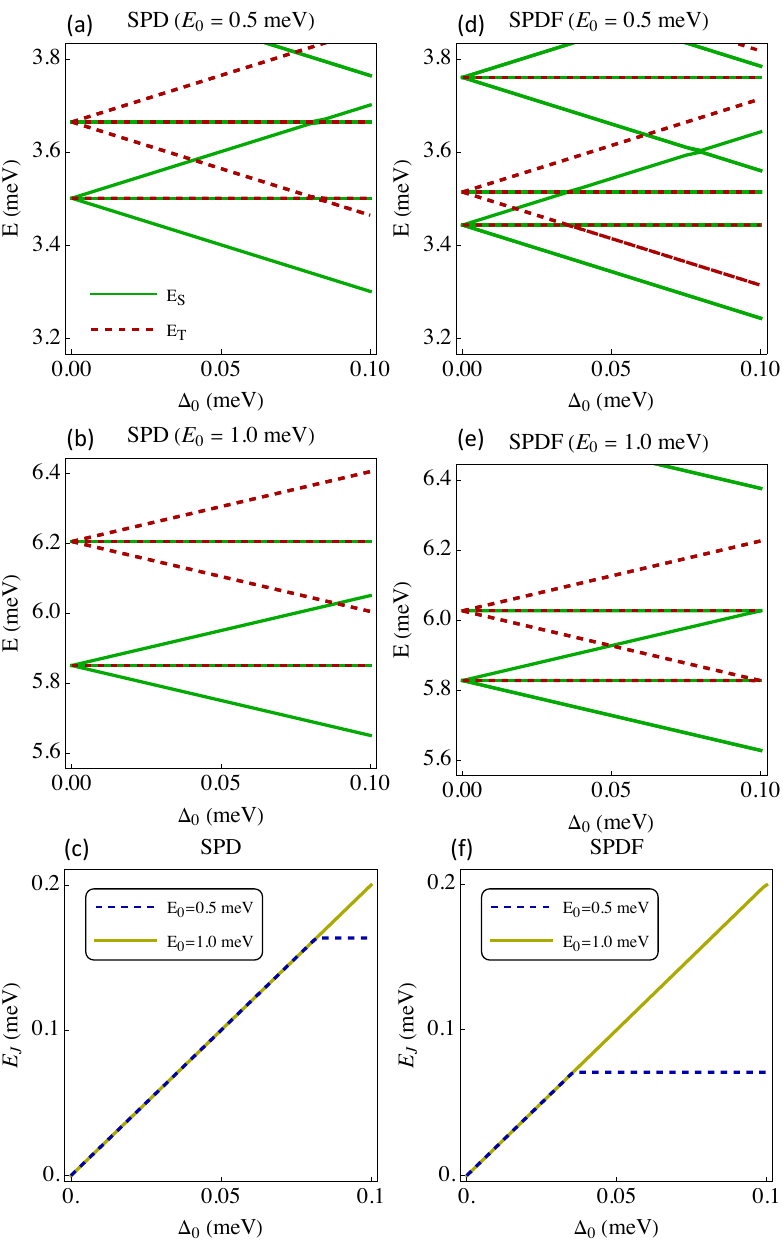}
	\caption{\textbf{Finite valley splitting energy spectrum:} (Color online) Energy levels of the few low-lying singlet (S$=$0) and triplet states (S$=$1) in a Si dot as a function of the valley splitting $|\Delta|$.  For this calculation we choose confinement energy $E_0=1.0$ and $0.5$ meV. The results of exchange energy with $|\Delta|$ with the $SPD$ and $SPDF$ are shown in the panels (c) and (e), respectively.}\label{fig_3}
\end{figure}

Under the assumption of a constant and uniform valley splitting, the single-electron energy spectrum in the quantum dot is split into two orthogonal valley manifolds, with identical structure and a constant shift (referred to as the valley splitting) relative to each other. Due to the orthogonality of the valley eigenstates, the two-electron spectrum is accordingly split into three orthogonal spaces, with both electrons in a single valley, or with one electron in each valley.  Figure \ref{fig_3} shows the two-electron spectrum calculated with single-electron basis set consisting of $SPD$ and $SPDF$ states, respectively, and two different confinement energies, of 0.5 meV and 1.0 meV, respectively.  With Coulomb matrix elements not affected by valley splitting, and Coulomb exchange integrals vanish when the electrons are in different valleys, the low-energy spectrum of two electrons in a single dot has only linear dependences on the valley splitting, via contributions from single-electron energy, as shown in Fig.~\ref{fig_3}.

%
%


When the valley splitting vanishes, the ground state is four-fold degenerate, containing three singlets and one triplet. Naively, the first excited level at zero valley splitting should be eight-fold degenerate, containing six triplets and two singlets. However, Coulomb interaction partly lifts this degeneracy, such that there are only three triplet and one singlet left in the first excited manifold. As valley splitting increases, singlets or triplets formed with electrons in different valleys remain unaffected, while singlets or triplets with both electrons in the same valleys vary linearly with valley splitting.  A more detailed analysis can be found in Appendix \ref{AP_2}.

In panels (a) and (d) of Fig.~\ref{fig_3}, when orbital excitation energy is $E_0=0.5$ meV, the energy gap between the orbital singlet and orbital triplet state is small. When valley splitting increases, there is a transition in the configuration of the ground triplet state, from a triplet with electrons in different valleys to one with both electrons in the lower valley. Such a transition is not observed at the stronger confinement energy of 1.0 meV, as the largest valley splitting of 0.1 meV is still not large enough to bring the different-valley triplet above the same-valley triplet in energy, as seen in panels (b) and (e).

The exchange splitting between the lowest-energy triplet and singlet states obtained with basis sets that include $SPD$ or $SPDF$ orbitals are shown in Fig.~\ref{fig_3}(c) and (f). At smaller valley splittings, the exchange splitting varies linearly with the valley splitting. In this region, the states are dominated by electrons in $s$-orbitals, with the singlet state having both electrons in the lower valley, and the triplet having one electron in each valley. Since the electrons are in different valley states, the triplet energy remains constant as valley splitting changes, while the singlet energy varies linearly with $2|\Delta_0|$. Beyond a certain value of valley splitting, 
the ground triplet crosses over to configurations with both electrons in the lower valley, leading to a flat region in the exchange splitting as both the ground singlet and triplet states have the same valley dependence.

\begin{figure}[t]
\minipage{0.45\textwidth}
  \includegraphics[width=8.0cm, height= 4.2cm]{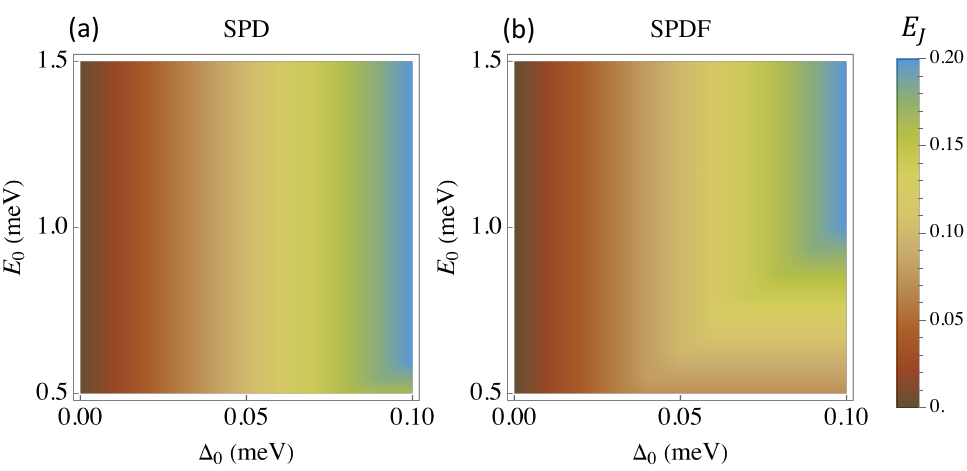}
\endminipage\hfill
   \caption{\textbf{$\bf{E_J \left(E_0,\,|\Delta_0|\right)}$ :} (Color Online) Density plot of the exchange interaction as a function of the confinement energy and the magnitude of valley-orbit coupling, with different sizes of the basis set $SPD$ and $SPDF$.}\label{fig_4}
\end{figure}

Figure~\ref{fig_4} shows how the singlet-triplet splitting changes with valley splitting and confinement energy.  It provides a complete picture of the transition between the two different types of triplets for both $SPD$ and $SPDF$ orbitals. Notably, when $F$ orbitals are included, the transition occurs at a larger confinement energy compared to when they are not included, showing the need to include the $F$ orbitals for a more accurate description of the energy spectrum. Figures~\ref{fig_3}(c) and (f) are essentially line cuts of this figure. 

Notice that a constant valley splitting and the same valley eigenstates mean that states from different valleys cannot couple, resulting in the so-called valley blockade\cite{culcer2012valley, yang2013spin, seedhouse2021pauli}. However, in a realistic situation, interface and alloy disorders are inevitably present, allowing different states to couple through valley-orbit coupling and lift the valley blockade, which we will study in the next section.

\section{Single-Electron Energy Levels of a Quantum Dot with Interface Roughness}\label{sec5}

In a realistic Si/SiGe heterostructure, both the magnitude and phase of the valley-orbit coupling are dependent on atomistic scale variations at or near the interface, such as alloy disorder in the barrier or atomic scale steps of the interface \cite{wuetz2021atomic, ercan2021strong, tariq2022impact, losert2023practical, klos2024atomistic, cakar2024toward}. Valley-orbit coupling is thus not decoupled from the electron orbitals anymore, since different charge distribution for each orbital would sample different interface disorders, which in turn leads to different valley-orbit coupling.  The most important consequence of such an orbital-dependent valley-orbit coupling is that we lose the valley-orthogonality among different orbital states, and the resulting selection rules and valley blockade.  We also face additional complexity in numerical calculations, in keeping track of valley splittings and phases for each orbital, as well as couplings between states \cite{buterakos2021spin}. 

The orbital-dependent valley-orbit couplings pose a significant challenge on how to vary them in a consistent manner: they should not be varied completely independently from each other, as they are all results of a given set of alloy disorder and interface roughness of a heterostructure.  For our calculation, which is based on the effective mass approximation, alloy disorder cannot be easily represented.  On the other hand, interface roughness can be modeled as a single atomic step on the interface. As such, we use a monolayer interface step, particularly its location, as a control knob to vary deterministically the valley-orbit couplings for all our orbitals in a consistent manner.  For convenience we choose a step along $y$ axis located at $x_0$, such that 
\begin{equation}
    \Delta^{ij}(x_0) = \langle D^i_z(\textbf{r})|H(x_0)|D^j_{\bar{z}}(\textbf{r})\rangle
\end{equation}
Here superscripts $i$ and $j$ represent the $i^{th}$ and $j^{th}$ orbitals for which we calculate the valley orbit couplings. The step height here is one monolayer, $d = a_0 / 4$, where $a_0$ is the lattice spacing of silicon. The presence of an interface step introduces a $2k_0d$ phase difference across the step for the electron wave function, which affects both the magnitude and phase of the valley-orbit coupling \cite{friesen2010theory, tariq2019}. 

For the case of a smooth interface (Appendix \ref{AP_1}), the off-diagonal valley orbit coupling terms are zero, while the diagonal terms are equal and can be tuned by the applied electric field along the growth direction. However, the presence of a step leads to the suppression in the magnitude of the valley-orbit coupling within same orbital for $i=j$ and introduce the valley orbit couplings in different orbitals for $i\neq j$  \cite{tariq2022impact}. The phase of the valley orbit coupling is sensitive to the step location and changes differently in each orbital due to the difference in the electron charge distribution. When the step is located at the center of the (symmetric) dot, the valley phase and magnitudes are the same in all the orbitals due to the mirror symmetry of the potential at this location.

\begin{figure}[h]
\includegraphics[width=\linewidth]{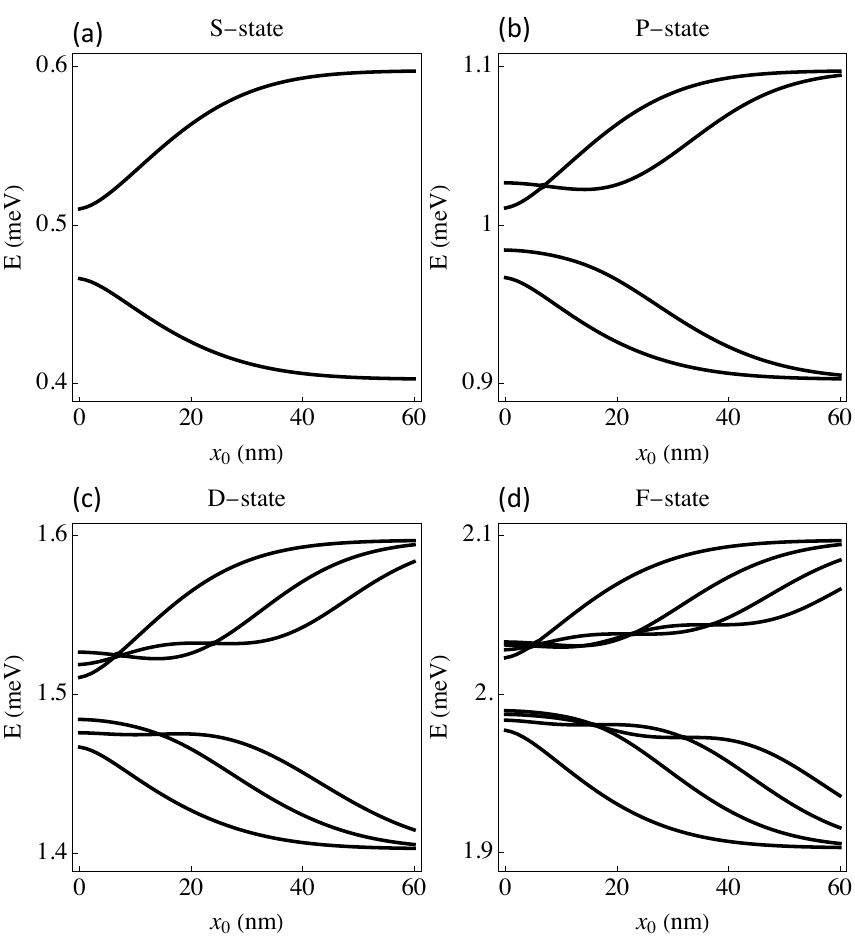}
   \caption{\textbf{One electron energy levels with step location:} Dependence of the energy levels of a single electron in different orbital states as a function of the step position. The height of the step is one monolayer and the confinement energy of quantum dot is 0.5 meV. The magnitude of the valley orbit coupling is 0.1 meV when the step is outside the quantum dot}\label{fig_5}
\end{figure}

In Fig.~\ref{fig_5}, we show that the location $x_0$ of the interface step significantly affects the single-electron energy spectrum in a quantum dot for different orbitals. It breaks the symmetry in the orbital states and splits each into two.  The step generally leads to a suppression of the valley splittings, with the strongest suppression happening when the step is located at the center of the quantum dot. Different orbitals (for example comparing S and P states) have slightly different $x_0$ dependence due to their particular charge distributions.  Lastly, the step also introduces coupling between different orbitals via valley-orbit couplings \cite{tariq2022impact}, causing further shifts in the energy levels.  This can be seen in the lack of symmetry around the original energy levels between any two valley-split orbitals (for example, for P-states, above and below 1 meV).
In short, the introduction of an interface step leads to orbital-dependent valley-orbit coupling, and mixing between orbital states. As a result, the valley selection rules that hold for a smooth interface, where transitions between different valley states are forbidden, are no longer valid.
%
%
%
%
%

\begin{figure}[h]
\includegraphics[width=\linewidth]{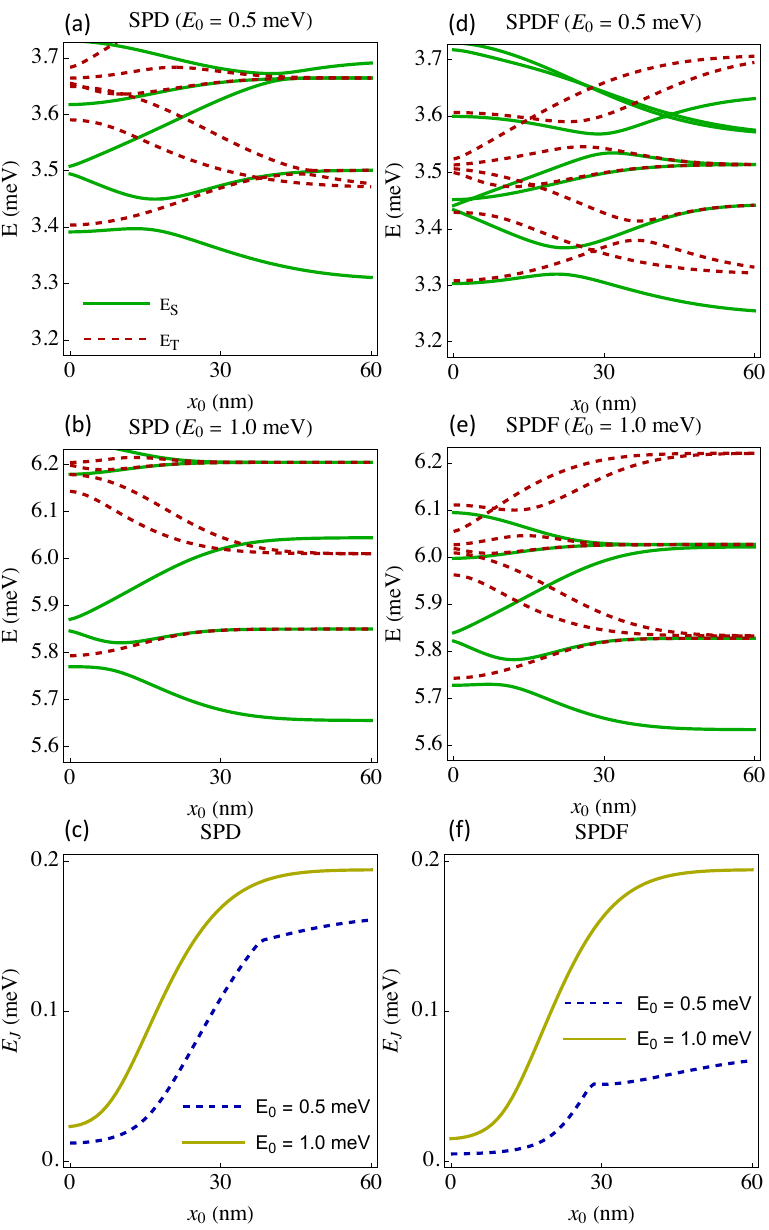}
   \caption{\textbf{Energy spectrum and exchange energy as functions of step locations:} The effect of the step position on the low-energy spectrum [panels (a), (b), (d), and (e)] and the exchange splitting [panels (c) and (f)] for a two-electron Si quantum dot in the presence of valley state.  Panels (a), (b) and (c) are obtained with basis states built from SPD single-electron orbitals, while panels (d), (e), and (f) are based on SPDF orbitals.}\label{fig_6}
\end{figure}

\section{Low-Energy Two-Electron Spectrum with Realistic Valley-Orbit Couplings}\label{sec6}

Having introduced general valley-orbit coupling for a single electron in a Si quantum dot, we now examine the spectrum and states of two electrons.  Here a key difference from the single-electron situation is the introduction of Coulomb interaction between the two electrons. The presence of valley states, which are highly oscillatory Bloch states, has an important impact on Coulomb matrix elements: for two electrons in different valleys, the Coulomb exchange integral vanishes.  Consequently, the three major influences, single-electron orbital excitation energy, valley-orbit coupling, and Coulomb interaction, compete to determine which configurations, whether two-valley or single-valley ones, dominate the two-electron states.  In this Section we first show our results on the low-energy spectrum of a two-electron Si quantum dot, then analyze the composition of these eigenstates, with a particular focus on how valley-orbit coupling affects the most important configurations in these states. 

\subsection{Low-Energy Spectrum in the Presence of an Interface Step}

Figure \ref{fig_6} shows the energy spectrum of a few low-lying singlet (solid green) and triplet (dashed red) levels as a function of the interface step location. We compare the results with $SPD$ and $SPDF$ orbital basis states at confinement energies of 0.5 and 1 meV. In each panel, the rightmost point ($x_0=60$ nm) corresponds to the case where the step lies outside the dimension of the quantum dot, approaching the smooth interface limit. 
%
%
%

In all four top panels the energy of the ground singlet state decreases as the step moves away from the center of the quantum dot.  This is reasonable assuming that the composition of the ground singlet is dominated by the configurations with both electrons in the lower valley, especially the doubly occupied lower-valley $S$ state.  For these configurations their energies depend on valley-orbit coupling as $-2|\Delta_0|$, thus the larger the valley splitting, the lower the energy of the ground singlet.

In panels (d) and (e) [as well as (a) and (b)] of Figure \ref{fig_6} there is an anticrossing between the ground and first excited singlet when $x_0$ is within the dot (the precise location is related to confinement energy).  When the step is outside the dot (large $x_0$), we know the ground singlet should be dominated by double-occupied $S$ orbital in the lower valley (same-valley singlet), especially if the confinement energy is large, as shown in Fig.\ref{fig_2}.  When the interface step is present inside the quantum dot, it reduces valley-orbit coupling in magnitude, and lifts all valley selection rules.  As such, the anticrossing here seems to be an indication of competition between same-valley and two-valley singlet configurations in the ground singlet state.  

In contrast to the ground singlet, the ground triplet energy increases as the interface step moves away from the center of the quantum dot. 
Since in a triplet the two electrons cannot occupy the same orbital state, it is usually believed that the ground triplet should be dominated by the configuration where the electrons are in the $S$ state of different valleys.  However, if this is indeed the case, the triplet energy should not depend on the step location: the energies of the two valleys change with the step as $|\Delta^{SS}|$ and $-|\Delta^{SS}|$, so that the sum has no dependence on $\Delta^{SS}$.  This apparent contradiction can only be clarified when we examine the state composition more closely in the next subsection.

Another feature of the triplet states in the case of a smaller confinement energy (0.5 meV) is a crossing between the ground and first excited triplet when the step is near the edge of the quantum dot.  This crossing is absent for the larger confinement energy (1.0 meV).  There the low-energy triplets seem to approach each other as the step moves away from the dot, as shown in panels (d) and (e) in Fig.~\ref{fig_6}. Clearly, the changing valley-orbit coupling is the origin of these features.  A clear physical picture will emerge when we analyze the state compositions carefully in the next subsection.

The exchange splitting between the ground triplet and singlet states are shown in Fig.~\ref{fig_6} (c) and (f). Here we observe a decease in the exchange energy when the step lies within the dimensions of the quantum dot, with its value reaching the minimum when the step is at the center of the dot. This behavior indicates that the exchange spitting is closely related to the valley-orbit coupling, which is also suppressed when the step approaches the center of the quantum dot. In the case of weaker confinement energy (dashed blue curve), a kink appears because of the transition between different configurations of the ground triplet state. In contrast, for strong confinement, the transition is absent and the curve remains smooth.  

To better understand the behaviors we observe in the low-energy spectrum of the two-electron quantum dot, we analyze the state composition of the ground singlet and triplet in the following Section. 

\subsection{State Composition in the Presence of an Interface Step}

The composition of the two-electron eigenstates goes a long way in determining the two-electron spectrum and its dependence on heterostructure properties.  Here we are particularly interested in how valley-orbit interaction, tuned by the location of an interface step, affects the most important contributing configurations, which in turn lead to the level crossing and anticrossing we observe in the energy spectrum. The procedure used to determine the state composition is outlined in Appendix \ref{AP_3}.

In Fig.~\ref{fig_7} we show how the contributions of different configurations to the ground singlet state depend on the location $x_0$ of an interface step. Panels (a) and (c) present results including orbitals up to the $D$ and $F$ shells, respectively, for a confinement energy of 0.5 meV. Panels (b) and (d) show the results for a stronger confinement energy of 1 meV. Only configurations that make the largest contributions {\it in the case of a smooth interface} are shown here. 

Figure \ref{fig_7} clearly shows a transition in the dominant configurations, corresponding to the anticrossing between the ground and first excited singlet states in Fig.~\ref{fig_6}.  For large $x_0$, with the step outside the quantum dot such that the interface within the range of the quantum dot is ideal, the most important configurations are what we expect from the perspective of an ideal smooth interface, with three orthogonal valley manifolds ($--$, $-+$, and $++$).  For the ground singlet state, the main configuration should be the doubly occupied $S$ orbital in the lower valley (i.e. large $|c^{ss}_{--}|^2$).  With valley blockade, the rest of the contributing configurations also have both orbitals in the $-$ valley, all represented by solid lines.  For smaller $x_0$, on the other hand, with the step inside the quantum dot, valley blockade is lifted, and valley-orbit coupling is orbital dependent.  Take the example of $x_0 = 0$ in Panel (c).  Here contributing configurations include both $+-$ and $--$ (and $++$ configurations as well, though they are not plotted here) valley occupations.  Furthermore, the $+-$ valley configurations are now dominant, with the largest contribution from a single configuration being the one where both electrons are in the $S$ orbital, but in different valleys, represented by $|c^{ss}_{-+}|^2$.  In other words, for a larger quantum dot without valley blockade, two-valley configurations are energetically more favorable. 

With a perfectly smooth interface, the circular quantum dot system we consider has rotational symmetry around the $z$ axis, such that the total $L_z$ would be a good quantum number for the two-electron states.  Consequently, the ground singlet state, which in this case is dominated by the doubly-occupied $S$ orbital in the lower valley and thus has $L_z = 0$, would only consist of those configurations that have $L_z = 0$, such as $SD_0$ and $P_+P_-$ configurations, as indicated by the solid lines in Fig.~\ref{fig_7} for large $x_0$. However, when an interface step is present inside a quantum dot, the rotational symmetry around $z$ is  broken, such that $L_z$ is not a good quantum number anymore.  In this case the two-electron eigenstates (whether singlet or triplet) could consist of configurations with any $L_z$ value, which is clearly illustrated in all the panels in Fig.~\ref{fig_7}.

\begin{figure}[t]
\includegraphics[width=\linewidth]{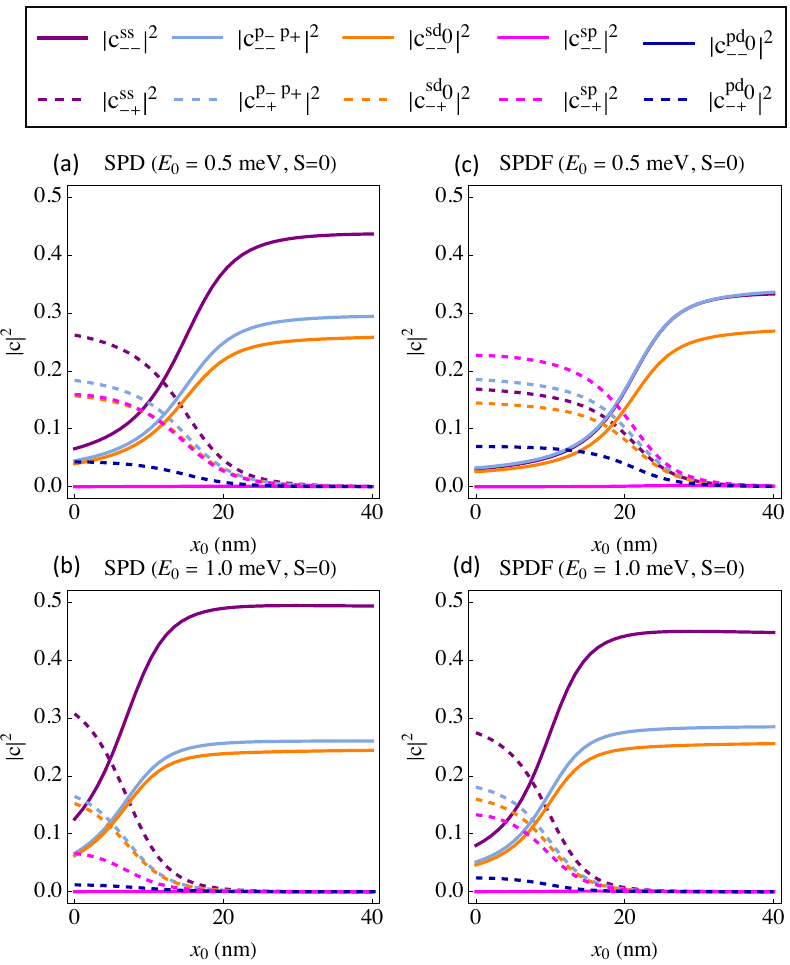}
   \caption{\textbf{State compositions of the ground singlets with step:} (Color online) The effects of the step on the state composition of the ground singlets with different number of orbitals and confinement energy. }\label{fig_7}
\end{figure}

\begin{figure}[h]
\includegraphics[width=\linewidth]{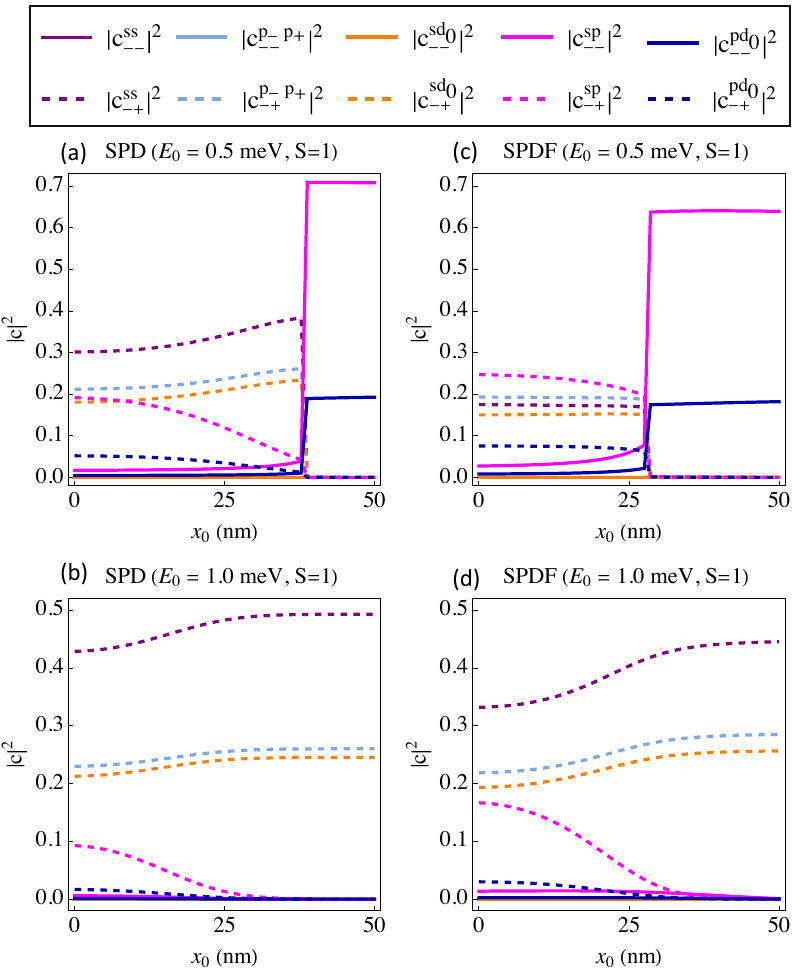}
   \caption{\textbf{State compositions of triplet with step:} (Color online) The impact of the step on the state composition of the ground triplet with different orbitals and confinement energy. The solid lines show the triplet states when one electron in the lower valley and other in the excited valley. Whereas, the dashed lines represents the triplets when both electrons in the lowest valley.}\label{fig_8}
\end{figure}

As shown in Fig.~\ref{fig_6} (a) and (d), there is a crossing between the ground and the first excited triplet states (red dashed lines), while in panels (b) and (e) such a crossing is absent.  This feature, or its absence, can again be traced to the state composition of the ground triplet state.  In Fig.~\ref{fig_8} we plot it as a function of the interface step position for both a large ($E_0 = 0.5$ meV, panels (a) and (c)) and a small dot ($E_0 = 1.0$ meV, panels (b) and (d)).  For the larger dot, with smaller orbital excitation energy, and the at the smooth interface limit (i.e. the step is outside the dot, or $x_0$ is large), the ground state turns out to be dominated by the configurations with both electrons in the lower valley, one in the $S$ orbital while the other in one of the $P$ orbitals.  By having both electrons in the lower valley, the Coulomb exchange integral (which vanishes if the two electrons are in orthogonal valleys) is finite, which helps lower the energy of such configurations.  On the other hand, when a step is inside the quantum dot, especially as it approaches the center of the dot, valley splitting decreases dramatically, such that the triplet with configurations where the two electrons are in different valleys (dashed lines in panels (a) and (c) of Fig.~\ref{fig_8}) becomes lower in energy.  In a smaller dot, with larger excitation energy, the gain from Coulomb exchange is never large enough to make the same-valley configurations the ground state.  Thus in Fig.~\ref{fig_8}(b) and (d) the ground state is always given by configurations with electrons in different valleys.

Similar to the case of the ground singlet state, the presence of the interface step inside the quantum dot breaks the rotational symmetry of the system, such that orbital angular momentum along the growth direction is not a good quantum number.  Consequently, for finite/small $x_0$, the configurations include both $L_z = 0$ ($ss$, $sd_0$, ...) and $L_z = 1$ ($sp$, $pd_0$) configurations.

\begin{figure}[h]
\includegraphics[width=\linewidth]{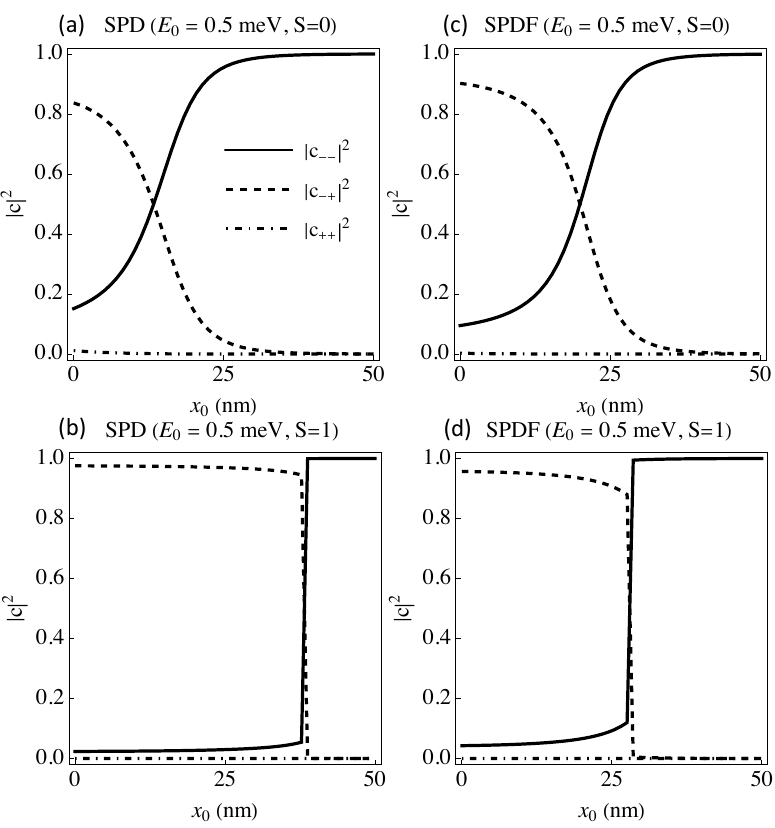}
   \caption{\textbf{Overall State Contribution:}  The state composition of the (a) ground singlet and (b) ground triplet states with an interface step for a dot confinement energy $0.5$ meV. Here the solid line show sum of all the contributions when both electrons in the lower valley state. Dashed line when one in the excited and other in the lower valley and the dot-dashed for both in the excited valley state. The major states contributing when the electrons are in different valleys for ground singlets and triplet are shown in (c) and (d), respectively. Note here $|c_{-+}^{sp}|^2=|c_{-+}^{sp_{-}}|^2+|c_{+-}^{sp_{-}}|^2+|c_{-+}^{sp_{+}}|^2+|c_{+-}^{sp_{+}}|^2$, i.e., all the possible contributions from different valley combinations of $s$ and $p$ orbitals. Similarly, $|c_{-+}^{ss}|^2=|c_{-+}^{ss}|^2+|c_{+-}^{ss}|^2$ and others states write in the same way.}\label{fig_9}
\end{figure}

In Fig.~\ref{fig_9}, we plot the overall contributions of electrons in the same-valley or different-valley configurations as a function of the step position. We find that the state composition of the ground triplet varies slowly with the step location, except for the transition from an orbital triplet to a valley triplet. In contrast, the ground singlet state composition varies smoothly and more dramatically while the step is inside the quantum dot, favoring different-valley configurations when the step is near the center of the quantum dot. 
We note that the contribution from $c_{++}$ is negligible, indicating that configurations with both electrons in the excited $+$ valley state are energetically unfavorable and thus play no significant role in the overall state composition.

Our results show that the state compositions of both the ground singlet and triplet states are sensitively dependent on the location of the interface step in a quantum dot due to how valley-orbit coupling (both its magnitude and phase) depends on the step location. The presence of an interface step breaks both valley blockade and rotational symmetry along the growth direction, allowing the ground singlet and triplet states to have a much larger variety of contributing two-electron configurations.
Most importantly, this conclusion can be readily generalized to the general Si/SiGe heterostructures where both interface roughness and alloy disorder are present. To reduce the impact of such disorder in valley-orbit coupling, the most straightforward measure is to reduce the quantum dot size and increase the orbital excitation energy, as we show in Appendix \ref{AP_4}.

\section{Effect of the Magnetic Field on Exchange Energy}\label{sec7}

When an external magnetic field is applied along the growth direction of a GaAs quantum dot, it induces singlet-triplet transitions in the two-electron ground state as the field strength grows, \cite{dineykhan1997two, nazmitdinov2009magnetic, baruffa2010spin} 
with the transition points in the parameter space depending on the confinement energy of the quantum dot.
Here we explore such magnetic field dependence in a two-electron silicon quantum dot, especially how valley-orbit coupling could impact the singlet-triplet transitions.  Below we first consider the orbital states by not including valley degree of freedom in the calculation to lay down the benchmarks for a quantum dot with Si parameters, then introduce the valley-orbit coupling. As mentioned in previous Sections, our calculations employ basis sets ranging from $SPD$ and up to $SPDFG$ levels. 

\subsection{Electron Spectrum in the Absence of Valley Degeneracy}

\begin{figure}[h]
\includegraphics[width=\linewidth]{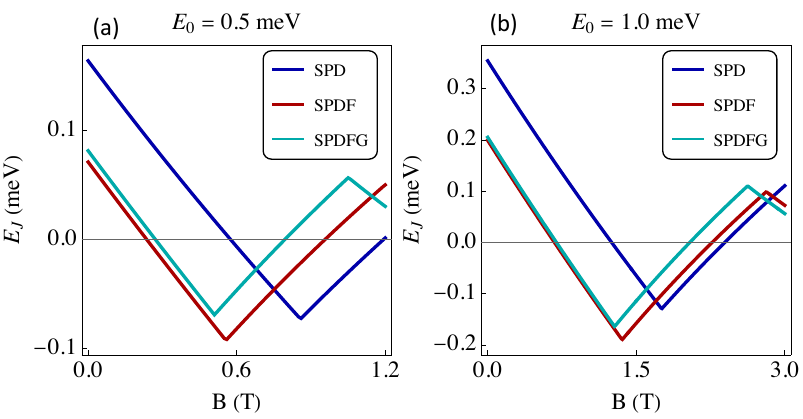}
   \caption{\textbf{Orbital exchange energy vs magnetic field:} The change in difference of the singlet triplet splitting with the magnetic field for different number of orbital states, excluding the valleys states. Panel (a) shows the result for smaller quantum dot of radius 20 nm corresponds to 1 meV confinement energy whereas (b) for 28 nm (0.5 meV) radius of the dot.}\label{fig_10}
\end{figure}

In Fig.~\ref{fig_10} we show the exchange splitting between the ground triplet and singlet states including up to $G$ orbitals in the basis states at the limit when valley is not in consideration. Here, regions where $E_J > 0$ represent the antiferromagnetic phase, when singlet is the ground state.  Conversely, when $E_J < 0$ a triplet is the ground state (without including Zeeman splitting). We show results up to $B=3$ T, where we observe a transition from a triplet ground state to a singlet with higher orbital momentum, i.e., $L_z=2$. Our results with $F$ and $G$ orbitals converge at lower magnetic fields, while at higher fields more excited orbitals need to be included in the basis set to achieve convergence. 

\begin{figure}[h]
\includegraphics[width=\linewidth]{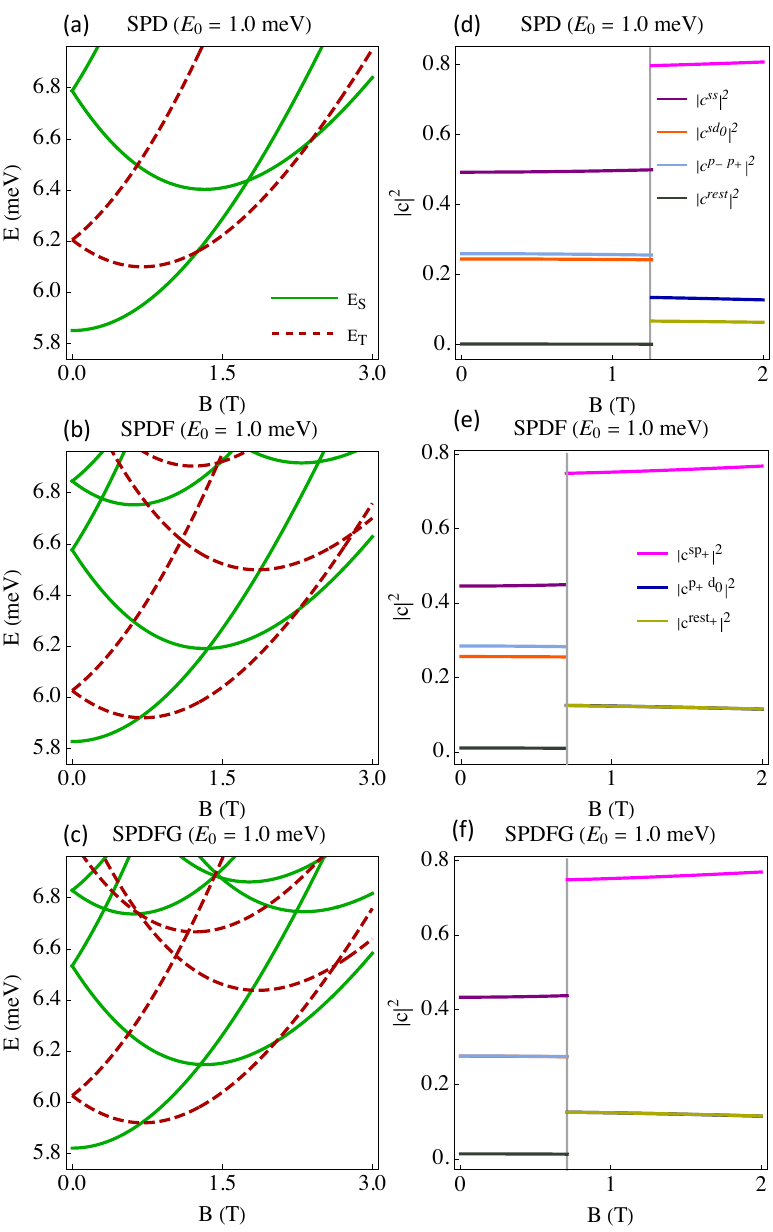}
   \caption{\textbf{The dependence of orbital singlets and triplets on magnetic field:} (Color online) The dependence of singlet ($S=0$) and triplet ($S=1$) states on the applied magnetic field is analyzed for different numbers of orbital states in the calculations. Valley contributions are excluded from this analysis. The energy levels are presented for (a) SPD, (b) SPDF, and (c) SPDFG basis sets, with the corresponding changes in state compositions shown in panels (d) to (f), respectively.}\label{fig_11}
\end{figure}

Figure \ref{fig_11} (a)-(c) show the energy levels of the low-lying singlet and triplet states as functions of the applied magnetic field in a quantum dot with a 1 meV confinement energy. Panels (a) and (b) clearly indicate that the ground singlet state at low field is barely affected by the inclusion of $F$ orbitals in the calculation. The excited singlets show a more marked decrease in energy with the inclusion of higher-energy orbitals, so that the crossing fields shift accordingly. On the other hand, all the triplet states, ground or excited, are strongly affected by the inclusion of the $F$ orbitals (but not nearly as much by the $G$ orbitals), with their energy levels decreasing by approximately 0.2 meV. Clearly, convergence for the triplet states requires at least the inclusion of $F$ orbitals in the CI calculation. 

The applied magnetic field reduces the effective confinement radius of the quantum dot and thus enhances the coulomb interaction between the two confined electrons. Figure \ref{fig_11} (d)-(f) present the state compositions of the ground state, which clearly show the first singlet-triplet transition. The state compositions of the singlet and triplet ground states barely change with the applied magnetic field near the transition. In other words, the effect of the magnetic field is mostly accounted for via the Fock-Darwin basis states, and the Fock-Darwin spectrum is essentially that of a simple harmonic oscillator with orbital magnetism, such that the relative weight of different configurations, especially those within the $L_z =0$ sector, do not have much of a magnetic field dependence. With an applied magnetic field, certain orbital degeneracies, such as the $L_z = \pm 1$ configurations represented by $sp_+$ and $sp_-$, are lifted as well. Last but not least, without considering spin-orbit interaction or valley-orbit coupling, the singlet-triplet transition here is sharp, showing a level crossing instead of an anti-crossing.

\subsection{Electron Spectrum in the Presence of Valley-Orbit Coupling}

We now investigate the impact of magnetic field on the two-electron spectrum in a silicon quantum dot in the presence of valley-orbit coupling. Our calculations here employ up to $F$ orbitals, and converge reasonably well at low magnetic fields. The confinement energy of the quantum dot is taken to be 1 meV. 

\begin{figure}[h]
\includegraphics[width=\linewidth]{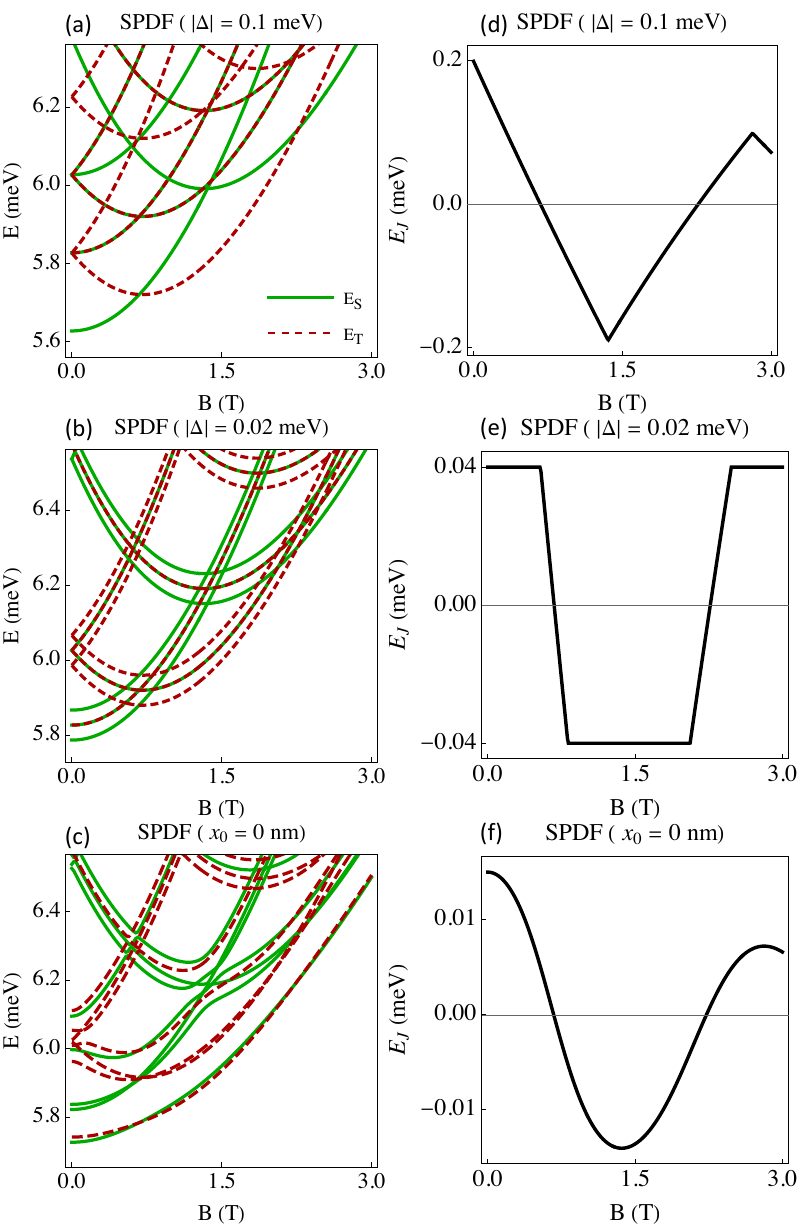}
   \caption{\textbf{Exchange splitting vs magnetic field with valleys:} (Color online) The transition between the singlet and triplet energy levels with the magnetic field using different number of orbital states including valley levels. The energy spectrum with constant valley splitting are shown in (a) 0.1 meV (b) 0.02 meV and in (c) for a fixed step located at the center of the dot. The corresponding exchange energy are shown from (d) to (f), respectively.}\label{fig_12}
\end{figure}

First, in Figs.~\ref{fig_12}(a) and (b), we consider two valley splitting magnitudes: 0.1 meV and 0.02 meV, respectively, under the ideal-interface assumption.  In other words we should have valley blockade in both these situations, and the two electron states can be divided into three categories, with both electrons in the lower valley, both in the higher valley, and one in each valley. For the larger valley splitting, the ground triplet is an orbital triplet (i.e. both electrons in the lower valley) rather than a valley triplet (one electron in each valley). The exchange splitting in this case is less than 0.2 meV and dominated by the orbital configurations in the lower valley for both singlet and triplet states. The magnetic field dependence of the exchange splitting arises purely from orbital magnetism determined by the cyclotron frequency, which is a linear dependence on the field, as shown in Figs.~\ref{fig_12}(d). In contrast, for the small valley splitting of 0.02 meV, the exchange splitting is 0.04 meV, and the orbital exchange is higher. Notably, in this small-valley-splitting situation, the exchange splitting is dominated by valley contributions, which has no magnetic field dependence, as shown in Figs.~\ref{fig_12}(e).

Similar to GaAs or single-valley Si situations discussed above, Figure \ref{fig_12}(a) and (b) show that as the applied magnetic field increases, the ground state transitions between the singlet and the triplet states. In both large- and small-valley-splitting situations, the singlet-triplet transition is caused by orbital magnetism (lowering of energy of $p_-$ state relative to $s$ state), though in the small valley splitting case additional states are present due to different valley configurations.

We have also examined a more realistic case when valley-blockade is lifted.  Specifically, we consider a special case where an interface step is located at the center of the Si quantum dot. 

Due to the central placement of the step, valley-orbit coupling is minimized in this case, while coupling between different valley states is also present. In Figs.~\ref{fig_12} (c) and (f) we observe a reduced exchange splitting and smooth dependence on the magnetic field, as both singlet and triplet states have contributions from all possible configurations due to the lack of symmetry.

\section{Conclusions}\label{sec8}

In this work, we employ the configuration interaction approach to investigate the low lying energy spectrum (with particular interest in the singlet-triplet exchange splitting) and state composition of a two-electron silicon quantum dot. Building upon orbital basis sets consisting of $SPD$ to $SPDFG$ single-electron orbitals, we demonstrate that excited orbitals, particularly up to those in $F$ shell, are essential for an accurate description of the few low-lying singlet and triplet states. Even in the absence of valley-orbit coupling, the ground singlet is not simply in the doubly occupied $SS$ configuration, and the ground triplet is not purely made of $SP$ configurations; instead, both contain substantial contributions from higher orbitals.  These results show that minimal orbital truncations are insufficient for reliable modeling of silicon spin qubits, as they can lead to substantial errors in exchange energies and wavefunction composition.

When a constant valley splitting is introduced, the two-electron spectrum separates into well-defined valley manifolds and exhibits valley blockade, with exchange splitting displays either linear or no dependence on the valley splitting depending on the orbital or valley configurations in the ground states. However, once realistic interface roughness (and/or other types of randomness, such as alloy disorder) is included through a monolayer step, valley-orbit coupling becomes orbital-dependent in both magnitude and phase, breaking valley orthogonality. As a result, valley selection rules are lifted. As a consequence, when the applied field or other knobs are varied, states usually anti-cross.  Wave function configurations would have smooth transitions in both singlet and triplet states (different spin configurations still do not mix, since we do not include spin-orbit interaction in our calculation), leading to strong reshaping of the singlet–triplet splitting. In particular, we show that the exchange splitting is minimized when the interface step is located near the center of the dot, where valley splitting is strongly suppressed. The ground singlet and triplet states can both transition between same-valley (orbital) and different-valley (valley) configurations depending on dot size and interface position, illustrating the competition among orbital excitation energy, Coulomb exchange, and valley-orbit coupling.

In the presence of a perpendicular magnetic field, and the absence of valley effects, singlet–triplet transitions are driven by orbital states and follow the Fock–Darwin spectrum. The state composition remains relatively stable near the singlet-triplet transitions. When valley–orbit coupling is present, the magnetic response depends strongly on the valley splitting. For large valley splitting, the exchange splitting is mainly orbital in origin and varies approximately linearly with magnetic field. For small valley splitting, valley contributions become dominant and the exchange splitting shows only weak field dependence. If valley blockade is lifted by interface disorder, the exchange splitting is reduced and changes smoothly with magnetic field due to strong configuration mixings.

Our results demonstrate that valley physics in silicon quantum dots cannot be treated as a simple correction to orbital models. Orbital excitation, Coulomb interaction, valley–orbit coupling, and atomic-scale interface roughness effects are connected and must be treated with accuracy to obtain a more precise picture of how singlet-triplet exchange splitting depends on system parameters and external perturbations. From a device perspective, the atomic structure of the interfaces (as well as the alloy disorder near the interfaces) emerges as a central design parameter for spin qubits in silicon since variations at these boundaries can significantly modify valley splitting and exchange splitting, thereby contributing to device-to-device variability. Strategies such as reducing dot size to increase orbital excitation energy and suppress valley mixing, alongside improving interface quality to stabilize valley splitting, provide clear pathways toward enhanced qubit reproducibility and coherence.

Overall, this work establishes a microscopic framework for understanding how valley–orbit coupling and realistic interface disorder reshape the two-electron spectrum in silicon quantum dots. These insights offer practical guidance for the modeling, interpretation, and engineering of scalable silicon-based quantum computing architectures.

\section*{Acknowledgements}
We thank support by US ARO through grants W911NF1710257 and W911NF2310018.

\section*{Competing interests}
The authors declare no financial or non-financial conflicts of interest.
\section*{Author contributions}
B.T. performed derivation and numerical calculation. B.T. and X.H. researched, analyzed, and prepared the manuscript.

\appendix
\begin{widetext}
    
\section{Valley Orbit Coupling}\label{AP_1}

For a single electron, the valley splitting at the $i^{th}$ orbital state is defined as follows:
\begin{equation}
    \Delta^{ii} = \langle D^i_z(\textbf{r})|H|D^i_{\bar{z}}(\textbf{r})\rangle.
\end{equation}
Here, $z$ and $\bar{z}$ refer to the two growth-direction valleys that contribute to the low-energy spectrum. For a smooth interface, the in-plane and out-of-plane directions of the Hamiltonian and wave functions are separable. Thus, the above expression can be simplified as follows:
\begin{equation}
    \Delta_0 = \int dz \; H_z  e^{2 ik_0z}|\phi_0(z)|^2 
\end{equation}

Where $\Delta_0\equiv \Delta^{ss}=...=\Delta^{d_{++}d_{++}}$ are all the diagonal elements. The above equation indicates that the value of the valley-orbit coupling depends on the potential along the $z$ direction, which is influenced by the barrier height and the electric field along the growth direction \cite{hosseinkhani2020electromagnetic, culcer2010interface, lima2023interface}.

\section{Energy Levels and State Composition with constant valley splitting}\label{AP_2}
Adding of a constant valley splitting to each orbital state doubles the energy levels for a single electron in a Si dot. This results in a larger set of singlet and triplet states in the two-electron case. Here, we drive the singlets and triplets formed by two electrons lie in two different orbital level if we include a constant valley splitting in them. As an example, we consider a system with one electron in the \( s \) orbital and another in the \( p \) orbital. For simplicity, we neglect the orbital degeneracy of the \( p \) state.  

In the absence of valley energy levels, the symmetric and antisymmetric orbital wavefunction in term of single electron wavefunction for one in $s$ and other in $p$ can be expressed as:
    \begin{eqnarray}
        \Psi_{S} ^{s,p} \left(\textbf{r}_1,\textbf{r}_2\right)&=&\frac{1}{\sqrt{2}}\left[\mathcal{F} ^{s} \left(\textbf{r}_1\right)\mathcal{F} ^{p} \left(\textbf{r}_2\right)+\mathcal{F} ^{p} \left(\textbf{r}_1\right) \mathcal{F} ^{s} \left(\textbf{r}_2\right)\right]\\
        \Psi_{A} ^{s,p} \left(\textbf{r}_1,\textbf{r}_2\right)&=&\frac{1}{\sqrt{2}}\left[\mathcal{F} ^{s} \left(\textbf{r}_1\right)\mathcal{F} ^{p} \left(\textbf{r}_2\right)-\mathcal{F} ^{p} \left(\textbf{r}_1\right)\mathcal{F} ^{s} \left(\textbf{r}_2\right)\right]
        \end{eqnarray}

The symmetric and antisymmetric wavefunctions, considering valley degrees of freedom (\( z \) and \( \bar{z} \)), are summarized in the following table \ref{tab_B1}:

\begin{table}[h]
    \centering
    \renewcommand{\arraystretch}{1.5}
    \begin{tabular}{|c|c|c|}
        \hline
        \textbf{Valley Configuration} & \textbf{Symmetric Wavefunction} & \textbf{Antisymmetric Wavefunction} \\ 
        \hline
        \( (\bar{z},\bar{z}) \) & \( \Psi_{S,\bar{z},\bar{z}}^{s,p} (\textbf{r}_1,\textbf{r}_2) = \frac{1}{\sqrt{2}} \left[ D_{\bar{z}}^{s} (\textbf{r}_1) D_{\bar{z}}^{p} (\textbf{r}_2) + D_{\bar{z}}^{p} (\textbf{r}_1) D_{\bar{z}}^{s} (\textbf{r}_2) \right] \)& \( \Psi_{A,\bar{z},\bar{z}}^{s,p} (\textbf{r}_1,\textbf{r}_2) = \frac{1}{\sqrt{2}} \left[ D_{\bar{z}}^{s} (\textbf{r}_1) D_{\bar{z}}^{p} (\textbf{r}_2) - D_{\bar{z}}^{p} (\textbf{r}_1) D_{\bar{z}}^{s} (\textbf{r}_2) \right] \) \\ 
        \hline
         \( (z,z) \) & \( \Psi_{S,z,z}^{s,p} (\textbf{r}_1,\textbf{r}_2) = \frac{1}{\sqrt{2}} \left[ D_{z}^{s} (\textbf{r}_1) D_{z}^{p} (\textbf{r}_2) + D_{z}^{p} (\textbf{r}_1) D_{z}^{s} (\textbf{r}_2) \right] \) & \( \Psi_{A,z,z}^{s,p} (\textbf{r}_1,\textbf{r}_2) = \frac{1}{\sqrt{2}} \left[ D_{z}^{s} (\textbf{r}_1) D_{z}^{p} (\textbf{r}_2) - D_{z}^{p} (\textbf{r}_1) D_{z}^{s} (\textbf{r}_2) \right] \) \\ 
        \hline
        \( (\bar{z},z) \) & \( \Psi_{S,\bar{z},z}^{s,p} (\textbf{r}_1,\textbf{r}_2) = \frac{1}{\sqrt{2}} \left[ D_{\bar{z}}^{s} (\textbf{r}_1) D_{z}^{p} (\textbf{r}_2) + D_{z}^{p} (\textbf{r}_1) D_{\bar{z}}^{s} (\textbf{r}_2) \right] \) & \( \Psi_{A,\bar{z},z}^{s,p} (\textbf{r}_1,\textbf{r}_2) = \frac{1}{\sqrt{2}} \left[ D_{\bar{z}}^{s} (\textbf{r}_1) D_{z}^{p} (\textbf{r}_2) - D_{z}^{p} (\textbf{r}_1) D_{\bar{z}}^{s} (\textbf{r}_2) \right] \) \\ 
        \hline
        \( (z,\bar{z}) \) & \( \Psi_{S,z,\bar{z}}^{s,p} (\textbf{r}_1,\textbf{r}_2) = \frac{1}{\sqrt{2}} \left[ D_{z}^{s} (\textbf{r}_1) D_{\bar{z}}^{p} (\textbf{r}_2) + D_{\bar{z}}^{p} (\textbf{r}_1) D_{z}^{s} (\textbf{r}_2) \right] \) & \( \Psi_{A,z,\bar{z}}^{s,p} (\textbf{r}_1,\textbf{r}_2) = \frac{1}{\sqrt{2}} \left[ D_{z}^{s} (\textbf{r}_1) D_{\bar{z}}^{p} (\textbf{r}_2) - D_{\bar{z}}^{p} (\textbf{r}_1) D_{z}^{s} (\textbf{r}_2) \right] \) \\ 
        \hline
    \end{tabular}
    \caption{Symmetric and antisymmetric wavefunctions in the presence of a constant valley states.}
    \label{tab_B1}
\end{table}
The Hamiltonian of the two electrons is:
\begin{equation}
    H_{2e}=H_{1e}(\textbf{r}_1)+H_{1e}(\textbf{r}_2)+H_{C}(\textbf{r}_1,\textbf{r}_2)\notag
\end{equation}
Using the eigen states we can construct the Matrix elements for the symmetric wave function and later write the results for anti-symmetric states too. Thus,
\[\langle \Psi_{S} ^{s,p} (\textbf{r}_1,\textbf{r}_2)|H_{2e}|\Psi_{S} ^{s,p} (\textbf{r}_1,\textbf{r}_2)\rangle=E_s+E_p+\langle \Psi_{S} ^{s,p} (\textbf{r}_1,\textbf{r}_2)|H_{C}|\Psi_{S} ^{s,p} (\textbf{r}_1,\textbf{r}_2)\rangle\] 
The contributions from terms with different valleys are very small due to the underlying orthogonal Bloch states and can easily be ignored such as, \(\langle D_{\bar{z}}^{i} (\textbf{r}_1) D_{\bar{z}}^{j} (\textbf{r}_2)|H_C|D_{\bar{z}}^{k} (\textbf{r}_1) D_{z}^{l} (\textbf{r}_2)\rangle=\langle D_{\bar{z}}^{i} (\textbf{r}_1) D_{\bar{z}}^{j} (\textbf{r}_2)|H_C|D_{z}^{k} (\textbf{r}_1) D_{z}^{l} (\textbf{r}_2)\rangle=0\) . The non-zero terms are

\begin{eqnarray}
    \langle  \Psi^{s,p}_{S,\bar{z},\bar{z}}|H_C| \Psi^{s,p}_{S,\bar{z},\bar{z}}\rangle &=& \langle D_{\bar{z}} ^{s} \left(\textbf{r}_1\right)D_{\bar{z}} ^{p} \left(\textbf{r}_2\right)|H_C| D_{\bar{z}} ^{s} \left(\textbf{r}_1\right)D_{\bar{z}} ^{p} \left(\textbf{r}_2\right)\rangle+ \langle D_{\bar{z}} ^{s} \left(\textbf{r}_1\right)D_{\bar{z}} ^{p} \left(\textbf{r}_2\right)|H_C| D_{\bar{z}} ^{p} \left(\textbf{r}_1\right) D_{\bar{z}} ^{s} \left(\textbf{r}_2\right)\rangle \notag\\
    \langle  \Psi^{s,p}_{S,z,z}|H_C| \Psi^{s,p}_{S,z,z}\rangle&=& \langle D_{z} ^{s} \left(\textbf{r}_1\right)D_{z} ^{p} \left(\textbf{r}_2\right)|H_C| D_{z} ^{s} \left(\textbf{r}_1\right)D_{z} ^{p} \left(\textbf{r}_2\right)\rangle+ \langle D_{z} ^{s} \left(\textbf{r}_1\right)D_{z} ^{p} \left(\textbf{r}_2\right)|H_C| D_{z} ^{p} \left(\textbf{r}_1\right) D_{z} ^{s} \left(\textbf{r}_2\right)\rangle \notag\\
    \langle  \Psi^{s,p}_{S,\bar{z},z}|H_C| \Psi^{s,p}_{S,\bar{z},z}\rangle&=& \langle D_{\bar{z}} ^{s} \left(\textbf{r}_1\right)D_{z} ^{p} \left(\textbf{r}_2\right)|H_C| D_{\bar{z}} ^{s} \left(\textbf{r}_1\right)D_{z} ^{p} \left(\textbf{r}_2\right)\rangle \notag\\
    \langle  \Psi^{s,p}_{S,z,\bar{z}}|H_C| \Psi^{s,p}_{S,z,\bar{z}}\rangle&=& \langle D_{z} ^{s} \left(\textbf{r}_1\right)D_{\bar{z}} ^{p} \left(\textbf{r}_2\right)|H_C| D_{z} ^{s} \left(\textbf{r}_1\right)D_{\bar{z}} ^{p} \left(\textbf{r}_2\right)\rangle \notag\\
    \langle  \Psi^{s,p}_{S,\bar{z},z}|H_C| \Psi^{s,p}_{S,z,\bar{z}}\rangle&=&  \langle D_{\bar{z}} ^{s} \left(\textbf{r}_1\right)D_{z} ^{p} \left(\textbf{r}_2\right)|H_C| D_{\bar{z}} ^{p} \left(\textbf{r}_1\right) D_{z} ^{s} \left(\textbf{r}_2\right)\rangle \notag\\
    \langle  \Psi^{s,p}_{S,z,\bar{z}}|H_C| \Psi^{s,p}_{S,\bar{z},\bar{z}}\rangle&=&  \langle D_{z} ^{s} \left(\textbf{r}_1\right)D_{\bar{z}} ^{p} \left(\textbf{r}_2\right)|H_C| D_{z} ^{p} \left(\textbf{r}_1\right) D_{\bar{z}} ^{s} \left(\textbf{r}_2\right)\rangle\notag
\end{eqnarray}
We define:
\begin{eqnarray}
u&=&\langle D_{\bar{z}} ^{s} \left(\textbf{r}_1\right)D_{\bar{z}} ^{p} \left(\textbf{r}_2\right)|H_C| D_{\bar{z}} ^{s} \left(\textbf{r}_1\right)D_{\bar{z}} ^{p} \left(\textbf{r}_2\right)\rangle=\langle D_{z} ^{s} \left(\textbf{r}_1\right)D_{z} ^{p} \left(\textbf{r}_2\right)|H_C| D_{z} ^{s} \left(\textbf{r}_1\right)D_{z} ^{p} \left(\textbf{r}_2\right)\rangle \notag\\
&=&\langle D_z ^{s} \left(\textbf{r}_1\right)D_z ^{p} \left(\textbf{r}_2\right)|H_C| D_{\bar{z}} ^{s} \left(\textbf{r}_1\right)D_z ^{p} \left(\textbf{r}_2\right)\rangle=\langle D_{z} ^{s} \left(\textbf{r}_1\right)D_{z} ^{p} \left(\textbf{r}_2\right)|H_C| D_{z} ^{s} \left(\textbf{r}_1\right)D_{z} ^{p} \left(\textbf{r}_2\right)\rangle \notag\\
k&=&\langle D_{\bar{z}} ^{s} \left(\textbf{r}_1\right)D_{\bar{z}} ^{p} \left(\textbf{r}_2\right)|H_C| D_{\bar{z}} ^{p} \left(\textbf{r}_1\right) D_{\bar{z}} ^{s} \left(\textbf{r}_2\right)\rangle=\langle D_{z} ^{s} \left(\textbf{r}_1\right)D_{z} ^{p} \left(\textbf{r}_2\right)|H_C| D_{z} ^{p} \left(\textbf{r}_1\right) D_{z} ^{s} \left(\textbf{r}_2\right)\rangle \notag\\
&=&\langle D_{\bar{z}} ^{s} \left(\textbf{r}_1\right)D_{z} ^{p} \left(\textbf{r}_2\right)|H_C| D_{\bar{z}} ^{p} \left(\textbf{r}_1\right) D_{z} ^{s} \left(\textbf{r}_2\right)\rangle=\langle D_{z} ^{s} \left(\textbf{r}_1\right)D_{\bar{z}} ^{p} \left(\textbf{r}_2\right)|H_C| D_{z} ^{p} \left(\textbf{r}_1\right) D_{\bar{z}} ^{s} \left(\textbf{r}_2\right)\rangle \notag
\end{eqnarray}

Thus the Matrix for symmetric and antisymmetric states can be written as,
\begin{equation}
    H_{S,2e}=E_s+E_p+\left[ {\begin{array}{cccc}
  u+k &0 & \Delta^* & \Delta^* \\
   0 & u+k & \Delta & \Delta \\
    \Delta & \Delta^* & u & k \\
    \Delta & \Delta^* & k & u \\
  \end{array} } \right],\;\;\;\;\;    H_{A,2e}=E_s+E_p+\left[ {\begin{array}{cccc}
  u-k &0 & \Delta^* & \Delta^* \\
   0 & u-k & \Delta & \Delta \\
    \Delta & \Delta^* & u & -k \\
    \Delta & \Delta^* & -k & u \\
  \end{array} } \right]
\end{equation}

In the above expression we assume that the valley orbit coupling in $s$ and $p$ orbitals are the same. Now, we consider two situations first zero valley splitting and then by including constant valley orbit couplings in the calculations.  
\subsection{$\Delta=0$}

The eigenvalues and eigen states of the above matrix is:

\begin{table}[h]
    \centering
    \renewcommand{\arraystretch}{1.5}
    \begin{tabular}{|c|c|}
        \hline
        \textbf{Eigenvalues} &  \textbf{Eigen states} \\ 
        \hline
        \( E_s+E_p+u+k  \) & \(\Psi^{s,p}_{S,\bar{z},\bar{z}}(\textbf{r}_1,\textbf{r}_2)\) \\ 
        \hline
        \( E_s+E_p+u+k \) & \(\Psi^{s,p}_{S,z,z}(\textbf{r}_1,\textbf{r}_2)\)\\
        \hline
        \( E_s+E_p+u+k \) & \( \frac{1}{\sqrt{2}} \left[\Psi^{s,p}_{S,\bar{z},z}(\textbf{r}_1,\textbf{r}_2)+\Psi^{s,p}_{S,z,\bar{z}}(\textbf{r}_1,\textbf{r}_2)\right] \) \\ 
        \hline
        \( E_s+E_p+u-k \) & \( \frac{1}{\sqrt{2}} \left[\Psi^{s,p}_{S,\bar{z},z}(\textbf{r}_1,\textbf{r}_2)-\Psi^{s,p}_{S,z,\bar{z}}(\textbf{r}_1,\textbf{r}_2)\right] \) \\ 
        \hline
    \end{tabular}
    \caption{Symmetric orbital wavefunctions and corresponding valley wavefunctions with $z$ and $\bar{z}$ basis sets for the case with $\Delta=0$.}
    \label{tab:B2}
\end{table}
In table \ref{tab:B2} $\Psi^{s,p}_{S,i,j}(\textbf{r}_1,\textbf{r}_2)$ and $\Psi^{s,p}_{A,i,j}(\textbf{r}_1,\textbf{r}_2)$ represent the symmetric and antisymmetric combination of valleys $i$ and $j$ and their combinations are stated in Table \ref{tab_B1}. In the above expression we can separate the orbital and valley contributions by writing, $D^{(i)}_z(\textbf{r}_1)\equiv \psi^{(i)}(\textbf{r}_1)\chi_z^{(1)},\;\;D^{(i)}_{\bar{z}}(\textbf{r}_1)\equiv \psi^{(i)}(\textbf{r}_1)\chi_{\bar{z}}^{(1)}$. Below we simplify the two electrons wave functions in to their respective orbitals and valley contributions as
\begin{eqnarray}
    &&\frac{1}{\sqrt{2}}\left[\Psi_{S,\bar{z},z} ^{s,p} \left(\textbf{r}_1,\textbf{r}_2\right)-\Psi_{S,z,\bar{z}} ^{s,p} \left(\textbf{r}_1,\textbf{r}_2\right)\right] \notag\\
    &=&\frac{1}{2}\left[\left\{D_{\bar{z}} ^{s}(\textbf{r}_1)D_{z} ^{p}(\textbf{r}_2)+D_{z} ^{p}(\textbf{r}_1)D_{\bar{z}} ^{s}(\textbf{r}_2)\right\}-\left\{D_{\bar{z}} ^{p}(\textbf{r}_1)D_{z} ^{s}(\textbf{r}_2)+D_{z} ^{s}(\textbf{r}_1)D_{\bar{z}} ^{p}(\textbf{r}_2)\right\}\right] \notag\\
    &=&\frac{1}{2}\left[\left\{\psi ^{s}(\textbf{r}_1) \psi^{p}(\textbf{r}_2)-\psi ^{p}(\textbf{r}_1)\psi ^{s}(\textbf{r}_2)\right\} \chi_z^{(1)}\chi_{\bar{z}}^{(2)}-\left\{\psi ^{s}(\textbf{r}_1) \psi^{p}(\textbf{r}_2)-\psi ^{p}(\textbf{r}_1)\psi ^{s}(\textbf{r}_2)\right\} \chi_{\bar{z}}^{(1)}\chi_z^{(2)} \right]\notag\\
    &=&\frac{1}{2}\left[\left\{\psi ^{s}(\textbf{r}_1) \psi^{p}(\textbf{r}_2)-\psi ^{p}(\textbf{r}_1)\psi ^{s}(\textbf{r}_2)\right\} \left[\chi_z^{(1)}\chi_{\bar{z}}^{(2)}- \chi_{\bar{z}}^{(1)}\chi_z^{(2)}\right] \right] \notag\\
    &=&\Psi_{A} ^{s,p} \left(\textbf{r}_1,\textbf{r}_2\right)\left[\frac{1}{\sqrt{2}}\left[\chi_z^{(1)}\chi_{\bar{z}}^{(2)}- \chi_{\bar{z}}^{(1)}\chi_z^{(2)}\right] \right] \notag
\end{eqnarray}
Using these results, we can apply the same procedure to the remaining states and express the wave functions in terms of their corresponding orbital and valley components for symmetric states as shown in Table \ref{tab:B3}.

\begin{table}[h]
    \centering
    \renewcommand{\arraystretch}{1.5}
    \begin{tabular}{|c|c|c|}
        \hline
        \textbf{Eigen states} & \textbf{Valley Contribution} & \textbf{Orbital Configuration} \\ 
        \hline
        \( E_s+E_p+u+k  \) & \(  \chi_{\bar{z}}^{(1)} \chi_{\bar{z}}^{(2)} \) & \( \frac{1}{\sqrt{2}} \left[ \psi^{s} (\textbf{r}_1) \psi^{p} (\textbf{r}_2) +\psi^{p} (\textbf{r}_1) \psi^{s} (\textbf{r}_2) \right] \) \\ 
        \hline
        \( E_s+E_p+u+k  \) & \( \chi_{z}^{(1)} \chi_{z}^{(2)} \) & \(  \frac{1}{\sqrt{2}} \left[ \psi^{s} (\textbf{r}_1) \psi^{p} (\textbf{r}_2) +\psi^{p} (\textbf{r}_1) \psi^{s} (\textbf{r}_2) \right] \) \\ 
        \hline
        \( E_s+E_p+u+k  \) & \( \frac{1}{\sqrt{2}} (\chi_{\bar{z}}^{(1)} \chi_{z}^{(2)} + \chi_{z}^{(1)} \chi_{\bar{z}}^{(2)}) \) & \(\frac{1}{\sqrt{2}} \left[ \psi^{s} (\textbf{r}_1) \psi^{p} (\textbf{r}_2) + \psi^{p} (\textbf{r}_1) \psi^{s} (\textbf{r}_2) \right] \) \\
        \hline
        \( E_s+E_p+u-k  \) & \( \frac{1}{\sqrt{2}} (\chi_{\bar{z}}^{(1)} \chi_{z}^{(2)} - \chi_{z}^{(1)} \chi_{\bar{z}}^{(2)}) \) & \(\frac{1}{\sqrt{2}} \left[ \psi^{s} (\textbf{r}_1) \psi^{p} (\textbf{r}_2) - \psi^{p} (\textbf{r}_1) \psi^{s} (\textbf{r}_2) \right] \) \\ 
        \hline
    \end{tabular}
    \caption{Symmetric wavefunctions breaks into their corresponding valley and orbitals wavefunctions.}
    \label{tab:B3}
\end{table}

In Table \ref{tab:B3}, we observe that the threefold degenerate levels with energy $E_s+E_p+u+k$ share the same orbital contribution but correspond to three distinct valley triplet states. In contrast, the final singlet state features an antisymmetric orbital contribution and a singlet valley configuration, with an energy level of $E_s+E_p+u-k$. This distinction explains why each singlet state splits into four: three states reside in the symmetric orbital configuration, while one corresponds to the antisymmetric orbital state.

Similarly, the contributions of the antisymmetric wave functions can be written in Table \ref{tab:B4}:
\begin{table}[h]
    \centering
    \renewcommand{\arraystretch}{1.5}
    \begin{tabular}{|c|c|c|}
        \hline
        \textbf{Eigen states} & \textbf{Valley Contribution} & \textbf{Orbital Configuration} \\ 
        \hline
        \( E_s+E_p+u-k  \) & \(  \chi_{\bar{z}}^{(1)} \chi_{\bar{z}}^{(2)} \) & \( \frac{1}{\sqrt{2}} \left[ \psi^{s} (\textbf{r}_1) \psi^{p} (\textbf{r}_2) -\psi^{p} (\textbf{r}_1) \psi^{s} (\textbf{r}_2) \right] \) \\ 
        \hline
        \( E_s+E_p+u-k  \) & \( \chi_{z}^{(1)} \chi_{z}^{(2)} \) & \(  \frac{1}{\sqrt{2}} \left[ \psi^{s} (\textbf{r}_1) \psi^{p} (\textbf{r}_2) -\psi^{p} (\textbf{r}_1) \psi^{s} (\textbf{r}_2) \right] \) \\ 
        \hline
        \( E_s+E_p+u-k  \) & \( \frac{1}{\sqrt{2}} (\chi_{\bar{z}}^{(1)} \chi_{z}^{(2)} + \chi_{z}^{(1)} \chi_{\bar{z}}^{(2)}) \) & \(\frac{1}{\sqrt{2}} \left[ \psi^{s} (\textbf{r}_1) \psi^{p} (\textbf{r}_2) - \psi^{p} (\textbf{r}_1) \psi^{s} (\textbf{r}_2) \right] \) \\
        \hline
        \( E_s+E_p+u+k  \) & \( \frac{1}{\sqrt{2}} (\chi_{\bar{z}}^{(1)} \chi_{z}^{(2)} - \chi_{z}^{(1)} \chi_{\bar{z}}^{(2)}) \) & \(\frac{1}{\sqrt{2}} \left[ \psi^{s} (\textbf{r}_1) \psi^{p} (\textbf{r}_2) + \psi^{p} (\textbf{r}_1) \psi^{s} (\textbf{r}_2) \right] \) \\ 
        \hline
    \end{tabular}
    \caption{Antisymmetric wavefunctions breaks into their corresponding valley and orbitals wavefunctions.}
    \label{tab:B4}
\end{table}
Here, we have again observed that the antisymmetric orbital parts multiplied with the three valley configurations and the symmetric orbital part which is higher in energy has the asymmetric valley configuration.
\subsection{$\Delta \neq 0$}
In the case of non-vanishing valley splitting we can replace our eigen states in terms of $\{-,+\}$ basis as,
\begin{eqnarray}
    D^i_\pm \left(\textbf{r}\right)&=&\frac{1}{\sqrt{2}}\left( D^i_z \left(\textbf{r}\right) \pm e^{ i \phi^i}D^i_{\bar{z}} \left(\textbf{r}\right)\right)\notag\\
    &=& \frac{1}{\sqrt{2}}\psi^i(r)\left( \chi^i_z \left(\textbf{r}\right) \pm e^{i \phi^i}\chi^i_{\bar{z}} \left(\textbf{r}\right)\right)\notag\\
    &=&\psi^i\left(\textbf{r}\right) \chi^i_\pm \left(\textbf{r}\right)\notag
\end{eqnarray}
where,
\begin{eqnarray}
    \chi^i_\pm \left(\textbf{r}\right)=\frac{1}{\sqrt{2}}\left( \chi^i_z \left(\textbf{r}\right) \pm e^{ i \phi^i}\chi^i_{\bar{z}} \left(\textbf{r}\right)\right)\notag
\end{eqnarray}
Since we have assumed the situation where the phase in $s$ and the $p$ state are the same then $\chi^s_\pm \left(\textbf{r}\right)=\chi^p_\pm \left(\textbf{r}\right)\equiv \chi_\pm \left(\textbf{r}\right)$. Thus, in Table \ref{tab:B3} and \ref{tab:B4} we replace the $\chi_{z,\bar{z}}$ with $\chi_{-,+}$ as to get the results for a constant valley splittings as given in Tables \ref{tab:B5} and \ref{tab:B6}:
\begin{table}[h]
    \centering
    \renewcommand{\arraystretch}{1.5}
    \begin{tabular}{|c|c|c|}
        \hline
        \textbf{Eigen states} & \textbf{Valley Contribution} & \textbf{Orbital Configuration} \\ 
        \hline
        \( E_s+E_p+u+k -2|\Delta| \) & \(  \chi_{-}^{(1)} \chi_{-}^{(2)} \) & \( \frac{1}{\sqrt{2}} \left[ \psi^{s} (\textbf{r}_1) \psi^{p} (\textbf{r}_2) +\psi^{p} (\textbf{r}_1) \psi^{s} (\textbf{r}_2) \right] \) \\ 
        \hline
        \( E_s+E_p+u+k +2|\Delta| \) & \( \chi_{+}^{(1)} \chi_{+}^{(2)}  \) & \(  \frac{1}{\sqrt{2}} \left[ \psi^{s} (\textbf{r}_1) \psi^{p} (\textbf{r}_2) +\psi^{p} (\textbf{r}_1) \psi^{s} (\textbf{r}_2) \right] \) \\ 
        \hline
        \( E_s+E_p+u+k  \) & \( \frac{1}{\sqrt{2}} \left(\chi_{-}^{(1)} \chi_{+}^{(2)} + \chi_{+}^{(1)} \chi_{-}^{(2)}\right) \) & \(\frac{1}{\sqrt{2}} \left[ \psi^{s} (\textbf{r}_1) \psi^{p} (\textbf{r}_2) + \psi^{p} (\textbf{r}_1) \psi^{s} (\textbf{r}_2) \right] \) \\
        \hline
        \( E_s+E_p+u-k  \) & \( \frac{1}{\sqrt{2}} \left(\chi_{-}^{(1)} \chi_{+}^{(2)} - \chi_{+}^{ (1)} \chi_{-}^{(2)}\right) \) & \(\frac{1}{\sqrt{2}} \left[ \psi^{s} (\textbf{r}_1) \psi^{p} (\textbf{r}_2) - \psi^{p} (\textbf{r}_1) \psi^{s} (\textbf{r}_2) \right] \) \\ 
        \hline
    \end{tabular}
    \caption{Symmetric wavefunctions breaks into their corresponding valley and orbitals wavefunctions for $\Delta\ne 0$.}
    \label{tab:B5}
\end{table}

\begin{table}[h]
    \centering
    \renewcommand{\arraystretch}{1.5}
    \begin{tabular}{|c|c|c|}
        \hline
        \textbf{Eigen states} & \textbf{Valley Contribution} & \textbf{Orbital Configuration} \\ 
        \hline
        \( E_s+E_p+u-k -2|\Delta| \) & \(  \chi_{-}^{(1)} \chi_{-}^{(2)} \) & \( \frac{1}{\sqrt{2}} \left[ \psi^{s} (\textbf{r}_1) \psi^{p} (\textbf{r}_2) -\psi^{p} (\textbf{r}_1) \psi^{s} (\textbf{r}_2) \right] \) \\ 
        \hline
        \( E_s+E_p+u-k +2|\Delta| \) & \( \chi_{+}^{(1)} \chi_{+}^{(2)}  \) & \(  \frac{1}{\sqrt{2}} \left[ \psi^{s} (\textbf{r}_1) \psi^{p} (\textbf{r}_2) -\psi^{p} (\textbf{r}_1) \psi^{s} (\textbf{r}_2) \right] \) \\ 
        \hline
        \( E_s+E_p+u-k  \) & \( \frac{1}{\sqrt{2}} \left(\chi_{-}^{(1)} \chi_{+}^{(2)} + \chi_{+}^{(1)} \chi_{-}^{(2)}\right) \) & \(\frac{1}{\sqrt{2}} \left[ \psi^{s} (\textbf{r}_1) \psi^{p} (\textbf{r}_2) - \psi^{p} (\textbf{r}_1) \psi^{s} (\textbf{r}_2) \right] \) \\
        \hline
        \( E_s+E_p+u+k  \) & \( \frac{1}{\sqrt{2}} \left(\chi_{-}^{(1)} \chi_{+}^{(2)} - \chi_{+}^{(1)} \chi_{-}^{(2)}\right) \) & \(\frac{1}{\sqrt{2}} \left[ \psi^{s} (\textbf{r}_1) \psi^{p} (\textbf{r}_2) + \psi^{p} (\textbf{r}_1) \psi^{s} (\textbf{r}_2) \right] \) \\ 
        \hline
    \end{tabular}
    \caption{Anti-symmetric wavefunctions breaks into their corresponding valley and orbitals wavefunctions for $\Delta\ne 0$.}
    \label{tab:B6}
\end{table}

The degeneracy is lifted in the presence of a constant valley splitting. The originally three-fold degenerate levels are now split by an energy difference of $2|\Delta|$, with two levels shifted above and below the original energy, while the third remains unaffected by the valley splitting. There are still degenerate energy levels of the symmetric and antisymmetric states where they both do not depends on the constant valley splitting. As a results, each orbital singlet or triplet state splits into three singlet or triplet levels and one additional triplet or singlet level, respectively. In the case of a rough interface, the number of energy levels remains the same; however, the valley splittings and the phases differ in each orbitals, resulting in a more general and complex energy spectrum.

\section{State Composition with step}\label{AP_3}
The easiest basis set to create the two-electrons wave function and the energy spectrum is the valley basis, $\{z,\bar{z}\}$. However, the general basis to define in term of the single electron basis, $\{-,+\}$, which is the  related to the valley basis as:

\begin{equation}
    D^i_\pm \left(\textbf{r}\right)=\frac{1}{\sqrt{2}}\left( D^i_z \left(\textbf{r}\right) \pm e^{ i \phi^i \left(\textbf{r}\right)}D^i_{\bar{z}} \left(\textbf{r}\right)\right)
\end{equation}
Where the superscript $^i$ represents the orbital state and the corresponding valley phase in that orbital as $ \phi^i$. Foe a smooth interface this value of $\phi^i$ is same for each orbital and thus for simplicity we can write it as $\phi_0$. Its value can be determined from Appendix A.

The symmetric combinations of the two electrons in the $z$ valley lying in different $i$ and $j$ orbitals can be written as:

\begin{equation}
    \Psi^{ij}_{zz}\left(\textbf{r}_1,\textbf{r}_2\right)=\frac{1}{\sqrt{2}}\left(D_z^i\left(\textbf{r}_1\right)D_z^j\left(\textbf{r}_2\right) + D_z^j\left(\textbf{r}_1\right) D_z^i \left(\textbf{r}_2\right) \right)
\end{equation}
In the above equation $\Psi^{ij}_{zz}$ depends on $(\boldsymbol{\textbf{r}_1}, \boldsymbol{\textbf{r}_2})$ with first $D$ has position $\boldsymbol{\textbf{r}_1}$ and second one is at $\boldsymbol{\textbf{r}_2}$. Similar equation can be written for the other combinations of the valleys. Thus, one can easily write the $++$ basis, e.g., into the $\{z,\bar{z}\}$ basis as:

\begin{eqnarray}
    \Psi^{ij}_{++}\left(\textbf{r}_1,\textbf{r}_2\right)&=&\frac{1}{\sqrt{2}}\left[D^i_+ \left(\textbf{r}_1\right)D^j_+ \left(\textbf{r}_2\right)+D^j_+ \left(\textbf{r}_1\right)D^i_+ \left(\textbf{r}_2\right)\right]\notag\\
    &=&\frac{1}{\sqrt{2}}\Bigg[ \frac{1}{2}\left( D^i_z \left(\textbf{r}_1\right) + e^{ i \phi_0}D^i_{\bar{z}} \left(\textbf{r}_1\right)\right)\left( D^j_z \left(\textbf{r}_2\right) + e^{ i \phi_0}D^j_{\bar{z}} \left(\textbf{r}_2\right)\right)\notag\\
    &&\;\;\;\;\;\;\;\; +\frac{1}{2}\left( D^j_z \left(\textbf{r}_2\right) + e^{ i \phi^j\left(\textbf{r}_2\right)}D^j_{\bar{z}} \left(\textbf{r}_2\right)\right)\left( D^i_z \left(\textbf{r}_1\right) + e^{ i \phi^i\left(\textbf{r}_1\right)}D^i_{\bar{z}} \left(\textbf{r}_1\right)\right)\Bigg]\notag\\
    &=&\frac{1}{2}\left[\Psi^{ij}_{zz}\left(\textbf{r}_1,\textbf{r}_2\right)+e^{i\phi_0}\Psi^{ij}_{z\bar{z}}\left(\textbf{r}_1,\textbf{r}_2\right)+e^{i\phi_0}\Psi^{ij}_{\bar{z}z}\left(\textbf{r}_1,\textbf{r}_2\right)+e^{2i\phi_0}\Psi^{ij}_{\bar{z}\bar{z}}\left(\textbf{r}_1,\textbf{r}_2\right)\right]
\end{eqnarray}
Similarly for other combinations we can write as

\begin{eqnarray}
    \Psi^{ij}_{--}\left(\textbf{r}_1,\textbf{r}_2\right) &=&\frac{1}{2}\left[\Psi^{ij}_{zz}\left(\textbf{r}_1,\textbf{r}_2\right)-e^{i\phi_0}\Psi^{ij}_{z\bar{z}}\left(\textbf{r}_1,\textbf{r}_2\right)-e^{i\phi_0}\Psi^{ij}_{\bar{z}z}\left(\textbf{r}_1,\textbf{r}_2\right)+e^{2i\phi_0}\Psi^{ij}_{\bar{z}\bar{z}}\left(\textbf{r}_1,\textbf{r}_2\right)\right]\notag \\
    \Psi^{ij}_{\pm\mp}\left(\textbf{r}_1,\textbf{r}_2\right) &=&\frac{1}{2}\left[\Psi^{ij}_{zz}\left(\textbf{r}_1,\textbf{r}_2\right)\pm e^{i\phi_0}\Psi^{ij}_{z\bar{z}}\left(\textbf{r}_1,\textbf{r}_2\right)\mp e^{i\phi_0}\Psi^{ij}_{\bar{z}z}\left(\textbf{r}_1,\textbf{r}_2\right)-e^{2i\phi_0}\Psi^{ij}_{\bar{z}\bar{z}}\left(\textbf{r}_1,\textbf{r}_2\right)\right]\notag
\end{eqnarray}
The probability of the electrons lying in the $ab$ valleys with electron in the $ij$ orbitals combinations is defined as,
\begin{equation}
    P^{ij}_{ab}=|c^{ij}_{ab}|^2=|\langle\Psi^{ij}_{ab}\left(\textbf{r}_1,\textbf{r}_2\right)|\Phi_{S,T}\left(\textbf{r}_1,\textbf{r}_2\right) \rangle|^2\
\end{equation}
%
Here, $\Phi_{S,T}$ denotes the wavefunction of the ground singlet or triplet state in the presence of a step and $a,b\in \{+,-\}$. It is important to note that $\Phi_{S,T}$ is expressed in the $z$ and $\bar{z}$ basis. To evaluate the state composition in the $\{a,b\}$basis, we write the $\Psi^{ij}_{ab}$ in terms of the valley phase of the electron in the $i^{\text{th}}$ and $j^{\text{th}}$ valley states. 


\section{Energy Spectrum with strong confinement energy ($3$ meV) quantum dot}\label{AP_4}
In the main text, we compared two types of quantum dots: one representing a larger dot (0.5 meV confinement energy) and the other a smaller dot (1 meV). With current technology, even smaller quantum dots can be fabricated. Here, we present the results and state compositions for an even smaller quantum dot with a size of approximately 11.6 nm, corresponding to a confinement energy of 3 meV \cite{zajac2016scalable, vandersypen2017interfacing, mills2019shuttling}. In this regime, the spacing between orbital energy levels increases, while the valley confinement energy remains unchanged. As a result, the influence of higher orbital states becomes even weaker.

\begin{figure}[h]
\includegraphics[width=\linewidth]{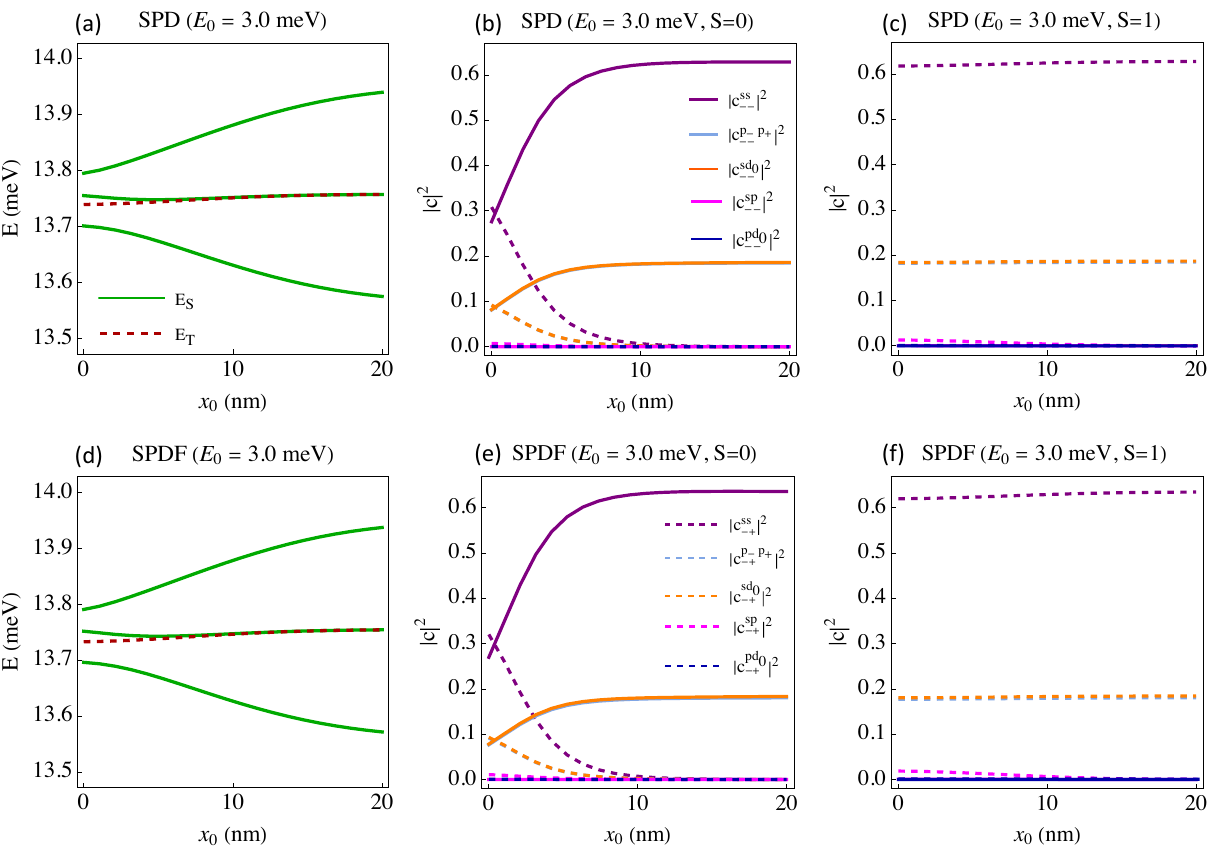}
   \caption{\textbf{Energy spectrum and state composition for confinement energy 3 meV:} (Color online) The low-lying singlet and triplet energy levels with the step location using orbital states up to (a) D and (d) F including valley levels. The state composition of the singlet and triplets with the states having high state compositions are shown in panels (b) and (c) for $SPD$ orbitals whereas, in panels (e) and (f) with $SPDF$ orbitals.}\label{fig_Ap1}
\end{figure}

We present the energy spectra for both the $SPD$ and $SPDF$ cases. The orbital levels are well separated for a strongly confined electrons in a quantum dot, resulting in only three singlet states and one triplet state being visible in both Fig. \ref{AP_1}(a) and (d). Due to the presence of a step within the quantum dot, the singlet states with energies $-2|\Delta|$ and $+2|\Delta|$ shift upward and downward, respectively, since both levels depends on the valley splitting and both suppressed with the inclusion of a step. The central line—corresponding to the singlet-triplet degeneracy observed at a smooth interface—shows no dependence on valley-orbit coupling under ideal situation like smooth interface. However, the introduction of the step breaks this symmetry, lifting the degeneracy between the two levels. Since the excited energy levels are well separated, the reduction in exchange energy is primarily driven by the suppression of valley splitting in the ground state.

The state composition of the singlets with both electrons occupying the same valley is suppressed in the presence of the step, while configurations with electrons in different valleys become more prominent. The contributions from the higher orbital states ($L=1$) are negligible, since the orbital energy levels are significantly larger than the valley splittings. Hence, in this case, most of the transitions occur within the orbital states $L=0$.

On the other hand, the composition of the triplet states does not vary with the step location. This is because the excited orbital energy levels lie much higher in energy, and therefore their influence on the ground-state composition is negligible. As a result, the variations observed in smaller quantum dots are not present in this case. Furthermore, for strong confinement, both the exchange energy and the state composition converge when orbital levels are included up to the $D$ orbital in the calculation. 
\end{widetext}

\bibliography{sample}

\end{document}